\documentclass[aps,prb,superscriptaddress,twocolumn,showpacs]{revtex4}
\usepackage{hyperref}
\usepackage{epsfig}
\usepackage{graphicx}
\usepackage{epstopdf}
\usepackage{tikz}
\usepackage{float}
\usepackage{amsmath}
\usepackage{amssymb}
\usepackage{bm}
\usepackage[normalem]{ulem}

\begin{document}
\title{Spin-sensitive interference due to Majorana state on\\ 
       interface between normal and superconducting leads}

\author{J. Bara\'nski}
\affiliation{Institute of Physics, Polish Academy of Sciences, 02-668 Warsaw, Poland} 

\author{A. Kobia\l ka}
\affiliation{Institute of Physics, M.\ Curie Sk\l odowska University, 20-031 Lublin, Poland}

\author{T. Doma\'nski}
\email{doman@kft.umcs.lublin.pl}
\affiliation{Institute of Physics, M.\ Curie Sk\l odowska University, 20-031 Lublin, Poland}

\date{\today}

\begin{abstract}
We investigate the subgap spectrum and transport properties of the quantum dot 
on interface between the metallic and superconducting leads and additionally
side-coupled to the edge of topological superconducting (TS) chain,  hosting the 
Majorana quasiparticle. Due to chiral nature of the Majorana states only one spin 
component of  the quantum dot electrons (say $\uparrow$) is directly affected, 
however the proximity induced on-dot pairing transmits its influence on the 
opposite spin as well. We investigate the unique interferometric patterns 
driven by the Majorana quasiparticle that are different for each spin component. 
We also address the spin-sensitive interplay with the Kondo effect manifested 
at the same zero-energy and we come to conclusion that quantum interferometry 
can unambiguously identify the Majorana quasiparticle. 
\end{abstract}  



\maketitle

\section{Introduction}
Many-body effects can generate in condensed matter systems a plethora 
of either bosonic (like phonons, magnons) or fermionic quasiparticles 
(e.g.\ polarons). Recently enormous activity has been devoted to very 
exotic type of quasiparticles, resembling the Majorana fermions 
\cite{Alicea-12,Flensberg-12,Stanescu-13,Beenakker-13,Franz-15} that are 
identical with their own antiparticles. Such emergent quasiparticles 
appear under specific conditions in the symmetry broken states 
\cite{Read-2000,Volovik-1999,Kitaev-2001} and their non-Abelian character  
makes them of interest for quantum computing and/or brand new  spintronic 
devices \cite{DasSarma-2016}. 

Realization of the Majorana quasiparticles has been predicted in various 
systems, for example in: vortices of superfluids \cite{Tewari-2007},
three-dimensional \cite{Fu-2008} or two-dimensional  \cite{Nilsson-2008} 
topological insulators coupled to superconductors, noncentrosymmetric 
superconductors \cite{Sato-2009}, electrostatic defects in topological 
superconductors \cite{Tworzydlo-2010}, $p$-wave superconductors \cite{Sau-2010},
the semiconducting \cite{Oreg-2010,Lutchyn-2010} or ferromagnetic \cite{Choy-2011} 
nanowires with the strong spin-orbit interaction  coupled to $s$-wave superconductors, 
Josephson junctions \cite{Aguado-2012}, ultracold atom systems \cite{ultracold},
and other. Experimental evidence for the Majorana
quasiparticles has been already reported by the tunneling spectroscopy using 
the Rashba nanowires coupled to the bulk $s$-wave superconductors 
\cite{Mourik-12,Yazdani-14,Kisiel-15,Franke-15}. Zero-bias enhancement 
of the differential conductance observed at the edges of such wires 
\cite{Mourik-12,Yazdani-14,Kisiel-15} has been interpreted as signature 
of the Majorana mode, but similar feature can be eventually assigned  
to disorder \cite{Liu-2012}, Kondo effect in a crossover from the doublet 
to singlet configurations \cite{Zitko-2015,Domanski-2016} or other effects
\cite{Rainis-2013}. 

\begin{figure}
\epsfxsize=6cm\centerline{\epsffile{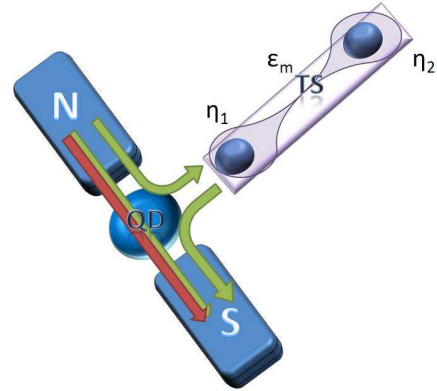}}

\vspace{-0.2cm}
\caption{Schematic illustration of the quantum dot (QD) laterally coupled to 
the metallic (N) and superconducting (S) electrodes and additionally 
hybridized with the Rashba nanowire, hosting the Majorana quasiparticles 
$\eta_{1}$ and $\eta_{2}$ at its edges. Green arrows indicate possible 
tunneling routes of $\uparrow$ electrons and the red arrow corresponds
to $\downarrow$ electrons.}
\label{schematics}
\end{figure}

For unambiguous identification of the Majorana quasiparticles there have 
been proposed several alternative methods, relying e.g.\ on optomechanical 
detection \cite{Chen-2014}, shot noise measurements \cite{Lutchyn-2015}, 
Josephson spectroscopy \cite{Flensberg-2016} using heterostructures 
comprising the quantum dot (QD) side-attached to the nanowire (see Fig.\ 
\ref{schematics}). In the case when both external leads are metallic 
it has been predicted reduction (by half) of the quantum dot conductance 
\cite{Baranger-2011}, suppression of the Seebeck coefficient (due to 
perfect particle-hole symmetry at the Fermi level) \cite{Leijnse-2014} 
and unique interferometric lineshapes \cite{Seridonio-2014,GongZhang-2014,
Jiang-2014,Desotti-2014,Stefanski-2015,Li-2015}.

T-shape setup, where QD is laterally coupled between the superconducting 
and metallic leads (Fig.\ \ref{schematics}) can reveal the fingerprints 
of Majorana fermions in the subgap spectrum \cite{Chirla-2016}. This 
configuration was already  addressed in the literature 
\cite{Chirla-2016,Gong-2014,Wang-2016} but the spin-sensitive transport 
properties have not been analyzed in detail. Charge transport can occur 
at low voltage via the Andreev scattering, when electrons from the normal 
lead are converted into the Cooper pairs of superconductor reflecting 
the holes back to the same metallic electrode. Such spin-selective Andreev 
spectroscopy has been suggested for probing the vortices in topological 
superconductors \cite{He-2014,Hu-2016} and it has recently provided 
evidence for the Majorana modes in Bi$_{2}$Te$_{3}$/NbSe$_{2}$ 
\cite{Sun-2016}. 

Two normal quantum dots arranged in the same T-shape configuration 
(as shown in Fig.\ \ref{schematics}) have been earlier studied by 
us \cite{Baranski-2011} and other groups 
\cite{Calle-2013,Nozaki-2014,Trocha-2014,Wojcik-2016}. These studies 
indicated that quantum interference effects are capable to probe an 
interplay between the electron paring (manifested by Andreev/Shiba states) 
and the strong correlations. Here we extend the previous analysis 
\cite{Baranski-2011}, by considering the influence of side-attached 
Majorana quasiparticle on the spin-resolved subgap spectrum of central 
QD and the transport properties. We show that Andreev transport would  
reveal interferometric lineshapes driven by the Majorana quasiparticles. 
Furthermore, we discuss how such interferometric features combine with 
the Kondo effect that is manifested at the same zero-energy. 

The paper is organized as follows. In Sec.\ II we formulate
the microscopic model and study interferometric effects appearing in 
a subgap spectrum of the uncorrelated QD. Next, in Sec.\ III, we analyze 
the correlation effect for the Kondo regime. Summary and conclusions are 
listed in Sec.\ IV. Some helpful technical details are presented in 
the Appendices.

\section{Formulation of the problem}

Due to chiral properties the ends of topological superconducting wire, 
that host a pair of Majorana fermions, are spin polarized 
\cite{Striclet-2012,Kjaergaard-2012,Shi-2016}. For this reason we 
assume that only one spin the central quantum dot (QD) in the T-shape 
configuration (Fig.\ \ref{schematics}) is directly coupled to the Majorana 
quasiparticle \cite{Vernek-2014}. When both electrodes are conducting 
the spin $\uparrow$ and $\downarrow$ transport channels would be independent, 
at least in absence of the correlations \cite{Vernek-2015}. This is however
no longer true, if one (or both) lead(s) is (are) superconducting, because 
of the proximity effect which mixes the particle with hole degrees of 
freedom \cite{Balatsky-2006,Domanski-2010}. In consequence, any physical 
process that engages electrons of a given spin simultaneously affects 
its opposite spin partner \cite{Golub-2015}. Such mechanism will prove 
to be important when considering the quantum interference driven 
by the side-coupled Majorana quasiparticle. 

The previous study \cite{Baranski-2011} indicated that electron pairing 
induced in the normal double quantum dot (DQD) on interface between the 
metallic and superconducting electrodes is characterized by two lineshapes: 
Fano-type resonance formed near the energy $\epsilon_{2}$ of the side-coupled 
quantum dot \cite{Zitko-2010} and anti-Fano structure appearing at $-\epsilon_{2}$ 
\cite{Baranski-2011}. These  features are detectable in 
the subgap Andreev conductance.
In the present study we  check whether similar effects appear when 
the central quantum dot is coupled to the Majorana quasiparticle, whose 
generic nature is related to only one spin  (say $\uparrow$).
For clarifying  the interferometric lineshapes appearing 
in the spectrum of QD  and the Andreev transport we briefly revisit 
the usual N-DQD-S heterostructure, imposing the spin polarized 
inter-dot hopping  (Appendix B). Such consideration provides useful interpretation 
of the spin-dependent Fano  and anti-Fano resonances.

\subsection{Microscopic model}

Tunneling structure, comprising the central QD embedded between the metallic 
and superconducting electrodes and side-coupled to the topological nanowire 
with the edge Majorana quasiparticles   (Fig.\ \ref{schematics}), can be 
described by the Anderson-type Hamiltonian
\begin{eqnarray}
H= H_{bath}+ \sum_{\beta=S,N}H_{T,\beta} +H_{QD} +H_{MQD}.
\label{HAnd}
\end{eqnarray}
The bath $H_{bath}=H_{N}+H_{S}$ consists of the metallic $H_{N}=\sum_{k, \sigma} 
\xi_{k N}C^{\dagger}_{k \sigma N}C_{k\sigma N}$ and superconducting  $H_{S}=
\sum_{k, \sigma} \xi_{kS}C^{\dagger}_{k \sigma S}C_{k \sigma S} - \sum_k 
( \Delta C^{\dagger}_{k\uparrow S}C^{\dagger}_{-k \downarrow S}+h.c.)$ 
reservoirs, where electron energies $\xi_{k\beta}$ are measured from the chemical 
potentials $\mu_{\beta}$.  The central  QD is described  by $H_{QD}=\sum_{\sigma} 
\epsilon d^{\dagger}_{\sigma} d_{\sigma}+Un_{\downarrow} n_{\uparrow}$, where 
$\epsilon$ denotes the energy level and $U$ stands for the repulsive interaction 
between opposite spin electrons.  QD is hybridized with the external reservoirs 
by $H_{T,\beta}=\sum_{k,\sigma}
(V_{k\beta}d^{\dagger}_{ \sigma} C_{k \sigma \beta}+h.c.)$, where 
$V_{k\beta}$ denote the matrix elements. 

Focusing on a subgap regime (i.e.\ energies $|\omega| \ll \Delta$) it 
has been shown  \cite{Bauer-07,Yamada-11,Baranski-2013} that the superconducting 
electrode  induces the static pairing. Its role can be thus played by 
the {\em proximized} quantum dot 
$H_{prox}=\sum_{\sigma} \epsilon d^{\dagger}_{\sigma} d_{\sigma}
+U n_{\downarrow}n_{\uparrow} - \frac{\Gamma_S}{2}(d_{\uparrow} 
d_{\downarrow}+d^{\dagger}_{\downarrow} d^{\dagger}_{\uparrow})$. 
This simplification is fairly acceptable for our considerations of 
the spin-dependent subgap spectrum and the Andreev spectroscopy. 
Low-energy theory of the Rashba nanowire can be expressed by
\cite{Lutchyn-2015}
\begin{eqnarray}
H_{MQD} = i\epsilon_m \eta_{1} \eta_{2} + \lambda ( d_{\uparrow} \eta_{1} 
+  \eta_{1} d^{\dagger}_{\uparrow})  ,
\label{Majorana_part}
\end{eqnarray}
where the operators $\eta_{i}=\eta_{i}^{\dagger}$ describe the edge states 
and $\epsilon_{m}$ accounts for their overlap. It is convenient to represent 
the exotic Majorana operators 
$\eta_1$, $\eta_2$ by the standard fermionic ones  \cite{Franz-15}
$\eta_{1}=\frac{1}{\sqrt{2}}(f+f^{\dagger})$,
$\eta_{2}=\frac{-i}{\sqrt{2}}(f+f^{\dagger})$. 
In this representation the term (\ref{Majorana_part}) takes the following form
\begin{eqnarray}
H_{MQD} = t_m (d^{\dagger}_{\uparrow} - d_{\uparrow}) 
( f + f^{\dagger} ) + \epsilon_m \left( f^{\dagger} f 
+ \frac{1}{2} \right) ,
\end{eqnarray}
where $t_{m}=\lambda/\sqrt{2}$.

\subsection{Scattering on Majorana quasiparticles}

Let us denote the particle  and hole Green's 
functions of the central QD coupled to the metallic and superconducting 
leads in absence of the Majorana quasiparticle by $\langle\langle d_{\sigma}; 
d^{\dagger}_{\sigma} \rangle\rangle \equiv a^{-1}$ and $\langle\langle 
d^{\dagger}_{\bar{\sigma}}; d_{\bar{\sigma}} \rangle\rangle\equiv b^{-1}$. 
For the uncorrelated case ($U=0$) these functions read (\ref{ORD_G}) 
\begin{eqnarray}
a & = & \omega-\epsilon+i\frac{\Gamma_N}{2} - 
\frac{(\Gamma_S/2)^2}{\omega+\epsilon+i\frac{\Gamma_N}{2}} ,
\label{a_symbol}
\\
b& = & \omega+\epsilon + i \frac{\Gamma_N}{2} 
- \frac{(\Gamma_S/2)^2}{\omega-\epsilon+i\frac{\Gamma_N}{2}}. 
\label{b_symbol}
\end{eqnarray} 
Similarly, we denote  the inverse particle and hole propagators 
of the isolated Majorana quasiparticle by $m \equiv  (\omega -\epsilon_m)$ 
and $n \equiv (\omega +\epsilon_m)$, respectively. 

Using the equation of motion approach (see Appendix A) 
we calculated  the matrix Green's function  
${\cal{G}}_{\sigma}(\omega)=\langle\langle \Psi_{\sigma} ; \Psi_{\sigma}
^{\dagger} \rangle \rangle$ defined in the matrix notation $\Psi_{\sigma}
= (d_{\sigma}, d_{\bar{\sigma}}^{\dagger}, f, f^{\dagger})$. For spin 
$\uparrow$ the Green's function reads
\begin{widetext}
\begin{eqnarray} 
{\cal{G}}_{\uparrow}(\omega) = \frac{1}{W}
\left( \begin{array}{cccc}  
bmn-2t_{m}^2\omega & -D(bmn-2t_{m}^2\omega) & bnt_{m} & bmt_{m}\\
-D(bmn-2t_{m}^2\omega) & \frac{1}{\omega-\epsilon+ i\Gamma_N/2}
+ D^2(bmn-2t_{m}^2\omega)& -Dbmt_{m}& -Dbnt_{m} \\
bnt_{m} & -Dbmt_{m} & abn - (a+b)t_{m}^2 & t_{m}^2(a+b)\\
bmt_{m} & -Dbnt_{m} & (a+b)t_{m}^2       & abm - t_{m}^2(a+b) 
\end{array}\right) , 
\label{Gr44}
\end{eqnarray} 
\end{widetext} 
where $W \equiv [abmn-2t_{m}^2\omega(a+b)]$ and $D  \equiv  (\Gamma_{S}/2)
/(\omega+\epsilon+i\Gamma_N/2)$. In the same way, we also determined the matrix 
Green's function ${\cal{G}}_{\downarrow}(\omega)$. Below we present explicitly 
${\cal{G}}_{\downarrow}^{(11)}(\omega)=\langle\langle d_{\downarrow} ; 
d^{\dagger}_{\downarrow} \rangle\rangle$ which yields the spectral function
of $\downarrow$ electrons. It takes the following form 
\begin{eqnarray}
{\cal{G}}_{\downarrow}^{(11)}(\omega) = G_N(\omega)+[\frac{\Gamma_S}{2} 
G_N(\omega)]^2\frac{amn-2t_{m}^2\omega}{abmn-2t_{m}^2\omega(a+b)}.
\label{Gdown}
\end{eqnarray}
 $G_N(\omega)=\langle\langle d_{\sigma}; d^{\dagger}_{\sigma} 
\rangle\rangle$ is the Green's function for the case when  QD is 
coupled only to the metallic lead (i.e. for $\Gamma_{S}=0=t_{m}$). 

Let us remark that differences between ${\cal{G}}_{\uparrow}(\omega)$ and 
${\cal{G}}_{\downarrow}(\omega)$ originate from the fact that only spin $\uparrow$
electrons are directly coupled to the Majorana mode. This difference vanishes for 
$t_{m} \rightarrow 0$ when ${\cal{G}}_{\sigma}^{11}$ reproduce the result 
\cite{Baranski-2013} obtained for dot coupled only to N and S electrodes 
$\lim_{t_m \rightarrow 0} {\cal{G}}_{\downarrow}^{11}={\cal{G}}_{\uparrow}^{11}=[\omega-
\epsilon+i\Gamma_N/2 -\frac{(\Gamma_{S}/2)^2}{\omega+\epsilon+i\Gamma_N /2}]^{-1}$. 
On the other hand, for $t_{m}\neq 0$  in absence of a superconducting electrode ($\Gamma_S=0$) the solution for spin $\downarrow$ electrons is identical with solution for QD coupled only to normal metal, regardless of the coupling strength to TS wire. This is because spin $\downarrow$  electrons are not affected by the side-coupled Majorana quasiparticle (so without electron pairing they do not `feel' any interference). This result clearly indicates that the 
interference patterns appearing in the spectrum of $\downarrow$ electrons originate solely 
from the pairing with electrons of the opposite spin.

\subsection{Deviation from usual Fano shape}

\begin{figure}
\epsfxsize=8cm\centerline{\epsffile{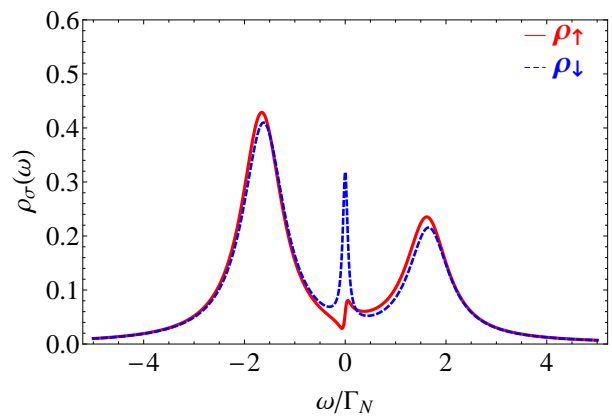}}
\caption{Spectral function $\rho_{\sigma}(\omega)$ of the central dot 
coupled to MQD for  $\epsilon_{m}=0$ using the model parameters 
$\Gamma_S=3\Gamma_N$, $\epsilon=-0.5\Gamma_N$,$t_m=0.3\Gamma_N$. 
We  notice that the interference 
pattern (at $\omega=0$) for $\uparrow$ electrons is different from 
the usual Fano shape.}
\label{Mdqd}
\end{figure}

In Fig.\ \ref{Mdqd} we illustrate  the spectral function $\rho_{\sigma}(\omega)$ 
of the central QD weakly coupled to the Majorana quasiparticle, in the case 
$\epsilon_{m}=0$. Interference pattern appearing  at $\omega=0$ in the spectral 
function $\rho_{\uparrow}(\omega)$ resembles a resonant lineshape. However, 
from a careful examination we clearly notice that it is not really the true Fano 
resonance (like the one for $\uparrow$ electrons shown in Fig.\ \ref{ordsep}). 
Such line-shape indicates, that electron waves resonantly scattered 
by the Majorana quasiparticle (to be regarded as half of the physical electron) 
change their phase only by the fraction of $\pi$ (which is typical value for 
scattering caused by the side-coupled ordinary quantum dots \cite{Zitko-2010}). 
Mechanism responsible for this fractional interferometric feature 
has the same origin as $4\pi$-periodicity of the Josephson junctions 
made of two `majoranized' superconducting wires 
\cite{Kitaev-2001,Lutchyn-2010,Aguado-2012,Fu-2009,Flensberg-2016}.

\begin{figure}
\epsfxsize=8cm\centerline{\epsffile{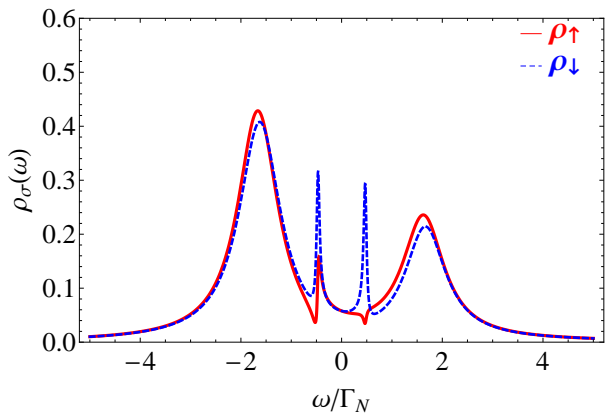}}
\caption{Spectrum of the  QD hybridized with the overlapping Majorana 
quasiparticles. Calculations have been done for $\epsilon_m=\Gamma_N$ using 
the same model parameters as in Fig.\ \ref{Mdqd}.}
\label{overlap}
\end{figure}

In Rashba nanowires of a finite length the Majorana quasiparticles partly 
overlap with one another, inducing some energy splitting between the edge 
modes  ($\epsilon_{m}\neq 0$). In Fig.\ \ref{overlap} we show the interference 
patterns obtained for $\epsilon_{m}=\Gamma_{N}$.
Spectrum of the spin $\uparrow$ electrons reveals two fractional Fano-like 
resonances (solid line in Fig.\ \ref{overlap}), whereas the spin $\downarrow$ 
electrons are characterized by two anti-Fano features (dashed lines in Fig.\ 
\ref{overlap}) at the same energies. Let us emphasize, that such behavior is 
qualitatively different from the results for T-shape heterojunction with the 
ordinary quantum dot (Fig.\ \ref{ordsep}). Interferometric effects 
could thus be  useful for detecting the Majorana quasiparticles in presence
of electron pairing.

\subsection{Evolution to  'molecular' region}

The interferometric Fano-like structures displayed in Figs \ref{Mdqd} and \ref{overlap}
occur when the central QD is very weakly coupled to the side-attached Majorana
quasiparticle. Upon increasing the interdot coupling $t_{m}$ the nanoscopic (QD 
and MQD) objects can be expected to develop some new spectroscopic signatures, 
characteristic for the entire `molecular' complex. 

\begin{figure}
\epsfxsize=7cm\centerline{\epsffile{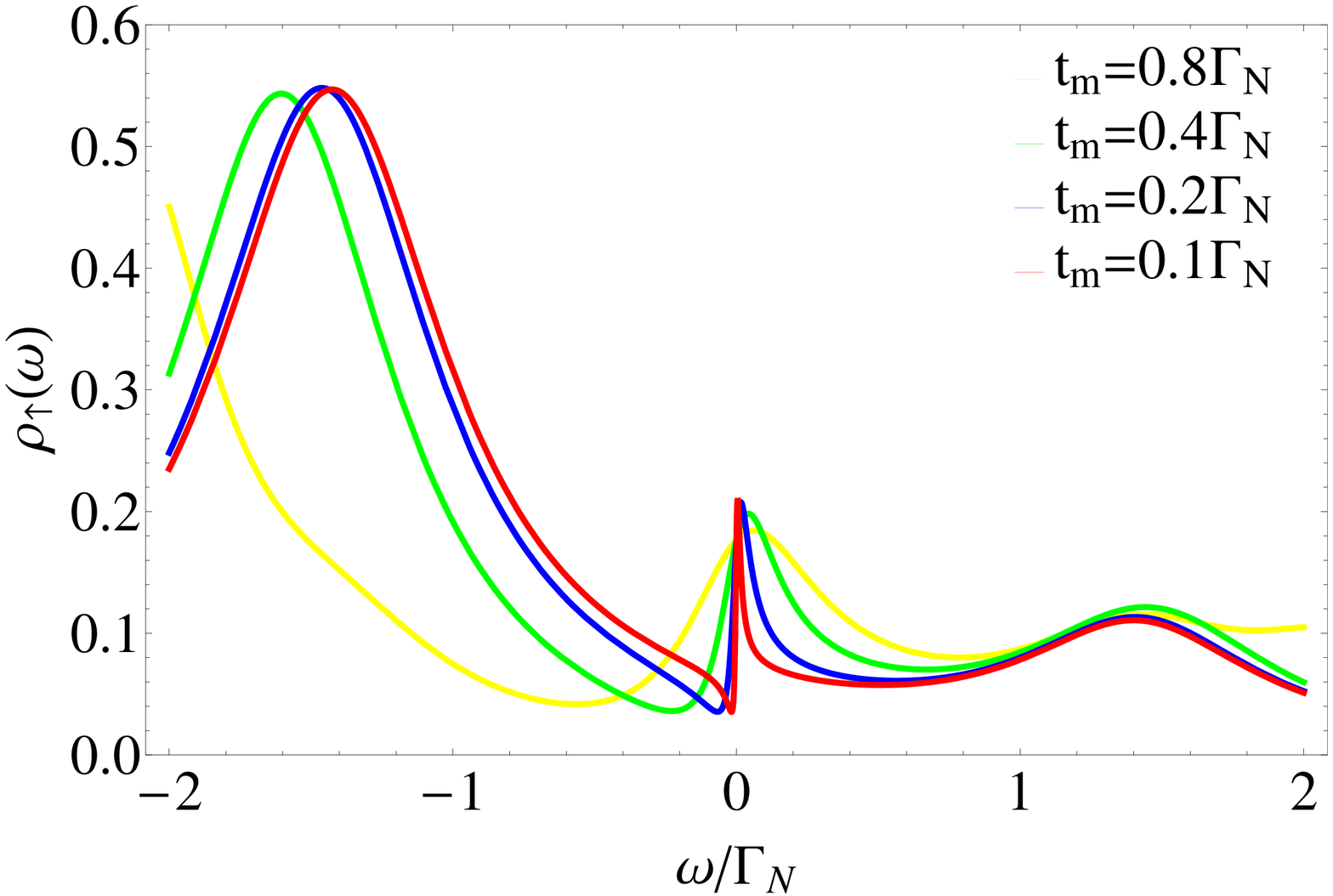}}
\epsfxsize=7cm\centerline{\epsffile{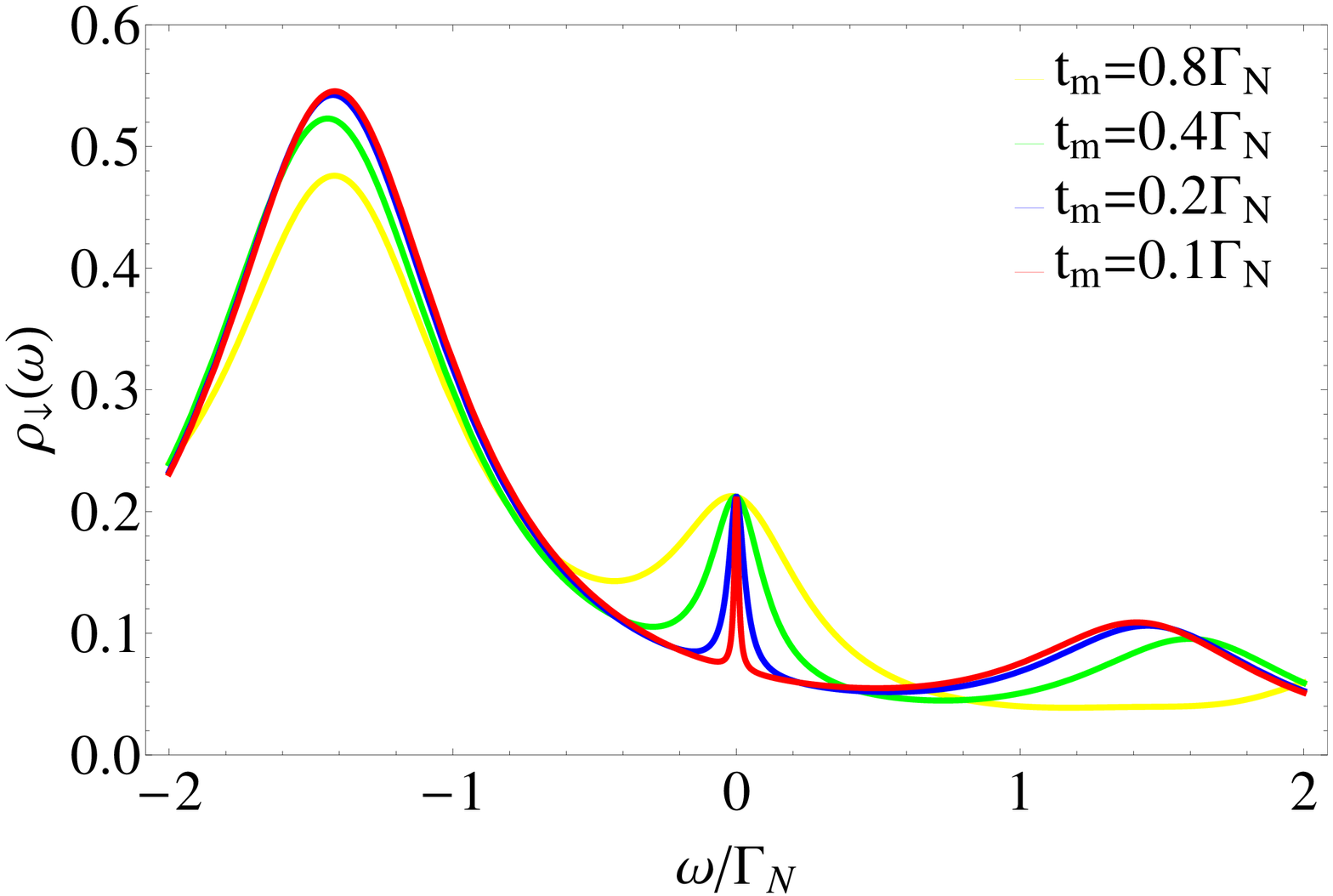}}
\caption{Spectral function $\rho_{\sigma}(\omega)$ of the central QD obtained 
for $\sigma= \uparrow$ (upper panel) and $\sigma= \downarrow$ (lower panel)
electrons, using $\Gamma_S=3\Gamma_N$, $\epsilon_{m}=0$ and various couplings 
$t_{m}$, as indicated.}
\label{molecular}
\end{figure}

Evolution of the spectral function $\rho_{\sigma}(\omega)$ vs $t_{m}$ is presented in 
Fig.\ \ref{molecular}. For increasing $t_{m}$ we observe that  
two Andreev peaks (originating from the mixed particle and hole degrees of 
freedom \cite{Baranski-2013}) and the Fano/anti-Fano lineshapes (caused by 
the Majorana QD) gradually change into the three-peak 
structure. In both spin components we clearly see emergence of the zero-energy 
peak at expense of reducing the spectral weight of the initial Andreev states. 
Formation of the zero-energy peak signifies a `leakage' of the Majorana 
quasiparticle into the central QD, in  analogy to what has been discussed 
in Ref.\ \cite{Vernek-2014}. In the present case such proximity induced 
zero-energy state affects both spin sectors, despite the fact that MQD 
is directly coupled only to the spin $\uparrow$ electrons.   

\subsection{Majorana fingerprints in Andreev spectroscopy}

Interference effects caused by the Majorana quasiparticle can be practically 
observed in our setup (Fig.\ \ref{schematics})  by measuring the tunneling
current under nonequilibrium conditions $\mu_{N}\neq \mu_{S}$. When applied 
voltage $ \mu_{N} - \mu_{S}\equiv eV $ is smaller in magnitude than $\Delta$ 
the charge current $I_{A}(V)=\sum_{i}I_{A i}(V)$ is contributed by spin 
$\uparrow$ ($i \equiv 1$) and spin $\downarrow$ ($i \equiv 2$) electrons.  
The spin-dependent Andreev currents can be expressed in  Landauer form 
\begin{eqnarray} 
I_{Ai}(V) = \frac{e}{h} \int \!\!  d\omega \; T_{Ai}(\omega)
\left[ f(\omega\!-\!eV)\!-\!f(\omega\!+\!eV)\right] ,
\label{I_A}
\end{eqnarray} 
where $f(x)=\left[ 1 + \mbox{\rm exp}(x/k_{B}T) \right]^{-1}$ is the 
Fermi distribution and transmittance for each spin sector 
%
\begin{eqnarray} 
T_{Ai}(\omega) = \Gamma_{N}^{2} \times \left\{ \begin{array}{cc}  
 \left| \langle\langle d_{\uparrow}; 
d_{\downarrow} \rangle\rangle
\right|^{2} & \mbox{\rm for} \hspace{0.2cm} i=1 \\
 \left| \langle\langle d_{\downarrow}; 
d_{\uparrow} \rangle\rangle
\right|^{2} & \mbox{\rm for} \hspace{0.2cm} i=2
\end{array} \right.
\label{T_A}
\end{eqnarray} 
describes a probability of converting the electron with spin $\sigma$
into the hole with spin $\bar{\sigma}$ in the metallic lead. 
The differential conductance $G_{A}(V)=dI_{A}(V)/dV$ is  
enhanced near the subgap (Andreev/Shiba) states \cite{Baranski-2013},  
but it is also sensitive to any other subgap features, 
including the quantum interference effects \cite{Baranski-2011}. 

\begin{figure}
\epsfxsize=8cm\centerline{\epsffile{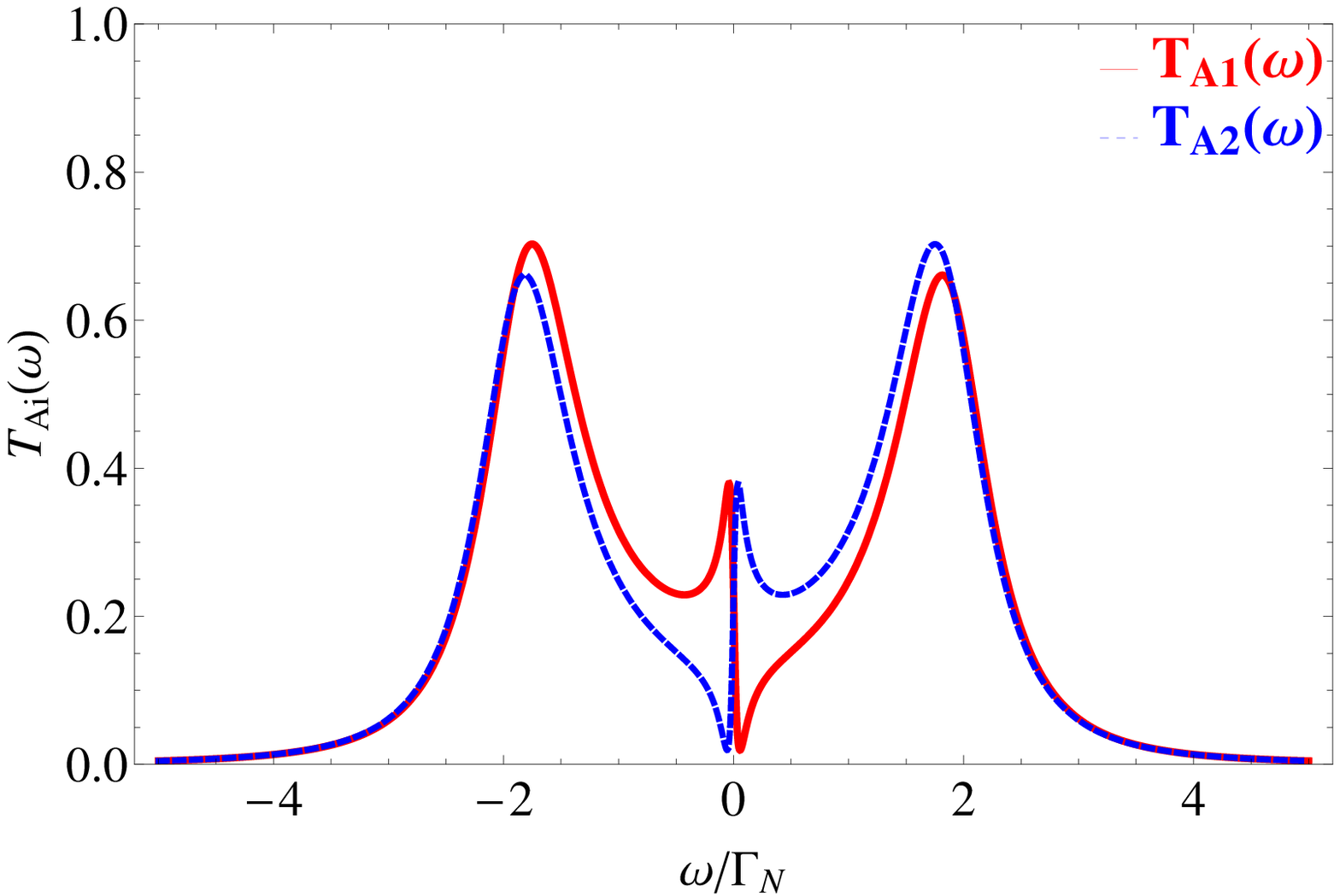}}
\epsfxsize=8cm\centerline{\epsffile{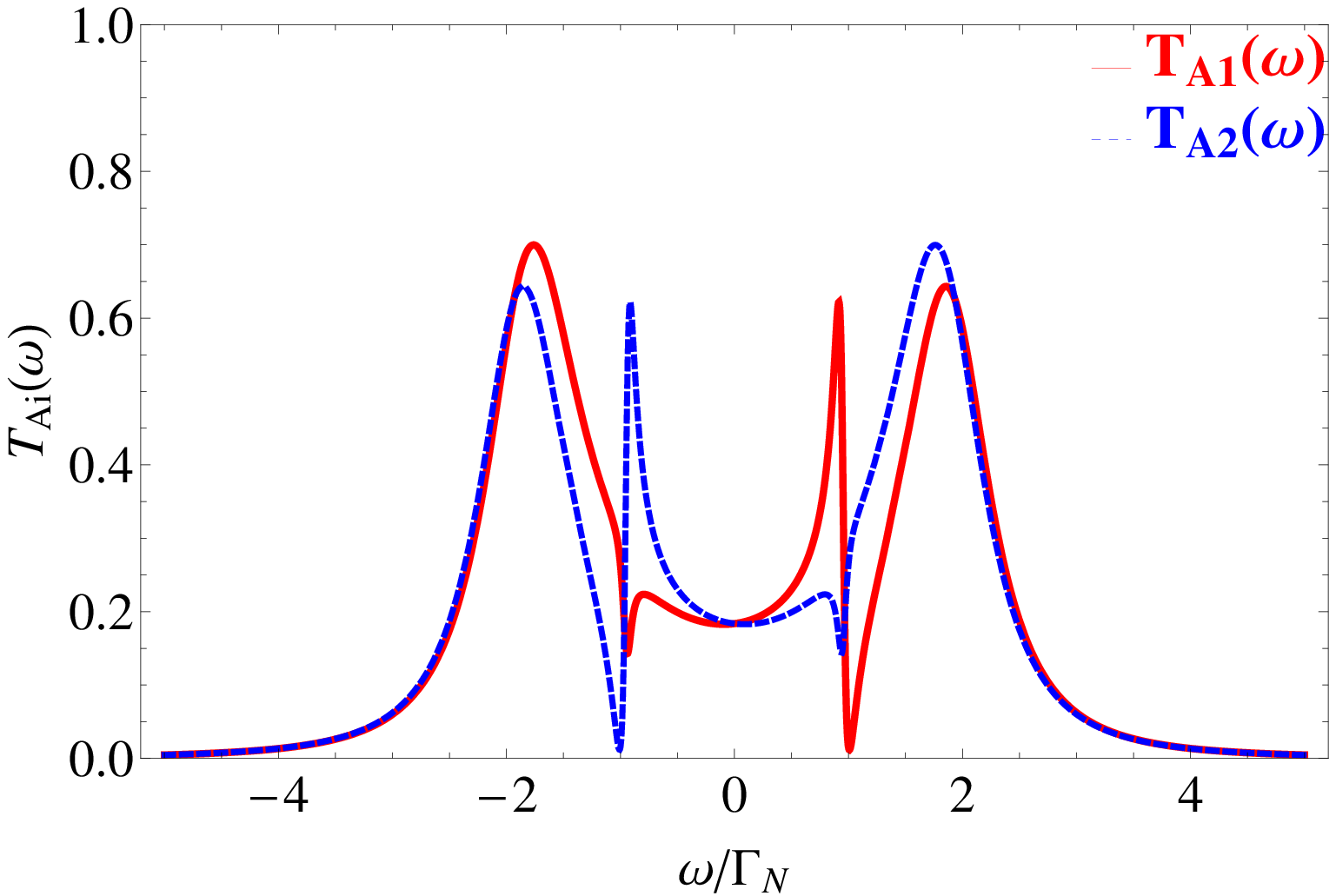}}
\caption{The spin-resolved Andreev transmittance of the quantum dot
strongly coupled to the superconducting lead $\Gamma_S=4\Gamma_N$ and weakly 
hybridized with the side-attached Majorana quasiparticle $t_{m}=0.3\Gamma_{N}$, 
where $\epsilon_{m}=0$ (upper panel) and $\epsilon_{m}=\Gamma_{N}$ (bottom panel). 
The red line refers to $\uparrow$ and the blue one to $\downarrow$ spin sectors.}
\label{Tra1}
\end{figure}

Fig.\ \ref{Tra1} shows the spin-resolved Andreev transmittance $T_{Ai}(\omega)$
obtained for $\epsilon_m=0$ (upper panel) and $\epsilon_m=\Gamma_{N}$ (bottom panel). 
In the first case we observe the fractional resonance appearing for each spin sector 
at zero-bias (although of opposite shapes). In the case $\epsilon_m \neq 0$ we notice 
two interferometric structures at $eV=\pm \epsilon_m$. For spin $\uparrow$ sector 
there appears the  pronounced resonance at $eV=\epsilon_m$ and another shallow 
structure at $eV=-\epsilon_m$. The Andreev transmittance of $\downarrow$ sector 
has an opposite shape, i.e.\ $T_{A2}(\omega)=T_{A1}(-\omega)$. 
The spin-resolved Andreev transport could thus estimate the overlap $\epsilon_m$
between the edge modes of TS nanowire. For $\epsilon_m=0$ such spectroscopy can 
distinguish the fractional interferometric lineshapes caused by the Majorana 
quasiparticle from the typical Fano/antiFano lineshapes due to the normal 
quantum dots (see Appendix B).

\section{Correlation effects}

Since the interferometric feature caused by the side-attached Majorana quasiparticle 
shows up at the Fermi level (for the case $\epsilon_m=0$) it is natural to inspect 
its relationship with the Kondo effect, whose signature (narrow peak) appears
at the same energy. The  many-body Kondo effect occurs at low temperatures due to 
the effective exchange interaction induced between the QD and normal lead (N) electrons.
Its subtle nature in a subgap regime has been addressed by variety of methods 
(see the recent discussion \cite{Zitko-2015,Domanski-2016} and other references cited therein). 

\subsection{Methodological details}

In the present context we shall treat the correlations using the decoupling
scheme for the Green's functions \cite{Baranski-2011} that proved to be satisfactory 
on a qualitative level \cite{Rodero-2011}. To account for the Kondo effect we 
start from the results obtained for the uncorrelated problem (section II) 
and proceed with approximations for the electron-electron interactions. In absence 
of the side-attached MQD, we again introduce the abbreviations for  particle 
$\langle \langle d_{\sigma}; d_{\sigma}^{\dagger} \rangle \rangle \equiv 
\tilde{a}^{-1}$ and hole $\langle \langle d_{\sigma}^{\dagger}; d_{\sigma}  
\rangle \rangle \equiv \tilde{b}^{-1}$ propagators. Following our
previous study \cite{Baranski-2013} of the single quantum dot  
(N-QD-S) setup we approximate these propagators by  
\begin{eqnarray}
\tilde{a} &=& \omega - \epsilon - \Sigma_{N}(\omega) + 
\frac{(\Gamma_S/2)^2}{\omega+\epsilon +[\Sigma_N(-\omega)]^{*}} ,
\label{new_a} \\
\tilde{b} &=& \omega + \epsilon +[\Sigma_N(-\omega)]^{*} + 
\frac{(\Gamma_S/2)^2}{\omega-\epsilon - \Sigma_N(\omega)} ,
\label{new_b}
\end{eqnarray}
where the selfenergy $\Sigma_N(\omega)$ accounts for the  coupling 
of QD to the normal lead, taking into account the interactions 
between electrons $Un_{\downarrow} n_{\uparrow}$.

We approximate $\Sigma_N(\omega)$ using the popular decoupling scheme 
for the Green's functions (discussed in Appendix B of Ref.\ 
\cite{Baranski-2011}) which yields
\begin{eqnarray}
\tilde{G}_N(\omega) &\equiv&  \frac{1}{\omega-\epsilon-\Sigma_N(\omega)}  
 \label{SigmaN} \\
&=&  \frac{\omega
-\epsilon-U(1-\langle n_{\sigma}\rangle) -\Sigma_3(\omega)]}
{[\omega-\epsilon][\omega-\epsilon-U-\Sigma_3(\omega)]
+i\frac{\Gamma_{N}}{2}U} ,
\nonumber
\end{eqnarray}
where
\begin{eqnarray}
\Sigma_{3}(\omega) = \sum_{k} |V_{k N}|^{2} 
 \left[ \frac{f(\xi_{k N})}{\omega\! - \! 
\xi_{k N} } + \frac{f(\xi_{k N})}{\omega\! -\! U \! - 
2 \varepsilon\!+\!\xi_{k N} } \right]  .
\label{sigmas} 
\end{eqnarray} 
Alternatively one can treat the correlation effects at the central 
quantum within more sophisticated methods \cite{Vernek-2015}.

Substituting the inverse Green's functions (\ref{new_a}, \ref{new_b}) with 
the  selfenergy $\Sigma_N(\omega)$  to the matrix Greens function 
(\ref{Gr44}) we  obtain
\begin{eqnarray}
{\cal{G}}_{\uparrow}^{11}(\omega)&=&\frac{\tilde{b}mn-2t_{m}^2\omega}{\tilde{W}} ,
\\
{\cal{G}}_{\downarrow}^{(11)}(\omega)&=&\tilde{G}_N(\omega)+\left[ 
\frac{\Gamma_S}{2} \tilde{G}_N(\omega) \right]^2
\frac{\tilde{a}mn-2t_{m}^2\omega}{\tilde{W}} ,
\nonumber \\ & &
\label{Gdown_bis}
\end{eqnarray}
where $\tilde{W}=\tilde{a}\tilde{b} m n- 2t_{m}^2\omega(\tilde{a}+\tilde{b})$.

In our setup the correlated quantum dot is connected to the superconducting reservoir,
which (by proximity effect) induces the on-dot electron pairing. On the other hand 
the repulsive Coulomb interactions disfavor any double occupancy, suppressing 
the local pairs. Even though the pairing and correlations are strongly antagonised 
one can find some regime of the model parameters, for which the Kondo physics 
coexists with the on-dot pairing \cite{Domanski-2016} (the latter is necessary
for activating the Andreev tunnelling that could probe the subgap states). 
This regime is particularly important if we want to confront the Kondo state 
with the interferometric structures due to side-attached Majorana quasiparticle.
 
Optimal conditions where the Kondo effect coexists with the on-dot pairing 
can be tuned by $\varepsilon$ (that controls QD occupancy) and 
the ratio between couplings to external the electrodes $\Gamma_S/\Gamma_N$ 
(that is crucial for the effective exchange potential \cite{Domanski-2016}).
For specific calculations we focus here on the strong Coulomb potential
$U=25\Gamma_{N}$ and choose $\epsilon=-2\Gamma_{N}$. We have checked 
that in such situation the Kondo effect coexists with the on-dot pairing for 
slightly asymmetric couplings $\Gamma_S\in (2\Gamma_N,6\Gamma_N)$. With this 
in mind, we thus fixed the ratio $\Gamma_S/\Gamma_N=4$. In absence of MQD 
(i.e. for N-QD-S configuration) the narrow Kondo peak at $\omega=0$ 
coexists then with the subgap Andreev quasiparticle peaks at $\omega \approx 
\pm \sqrt{\epsilon^{2}+(\Gamma_S/2)^{2}}$ whose broadening (inverse 
life-time) is proportional to $\Gamma_{N}$ \cite{Baranski-2013}.

\subsection{Majorana vs Kondo feature}

\begin{figure}[H]
\epsfxsize=7cm\centerline{\epsffile{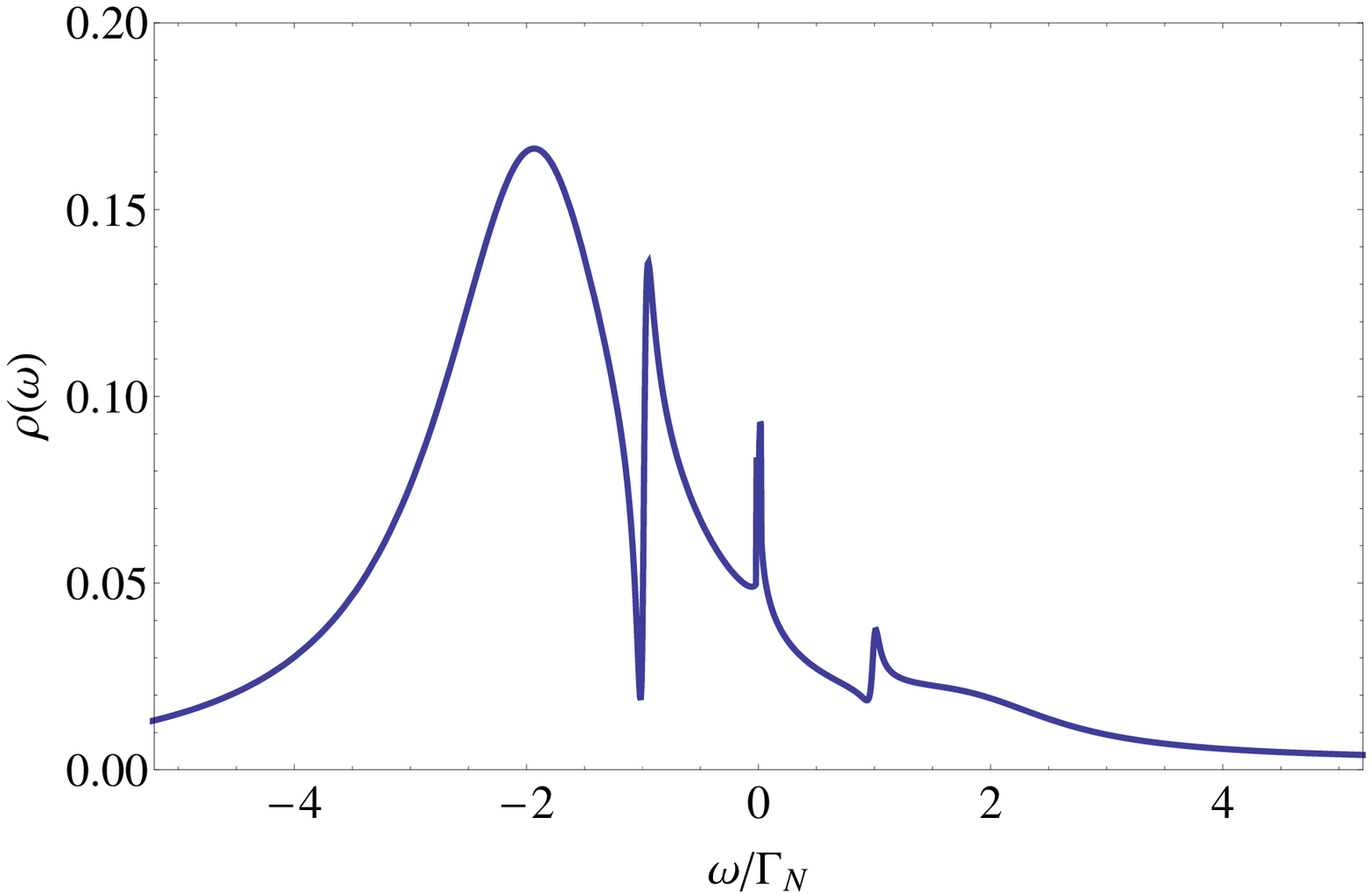}}
\epsfxsize=7cm\centerline{\epsffile{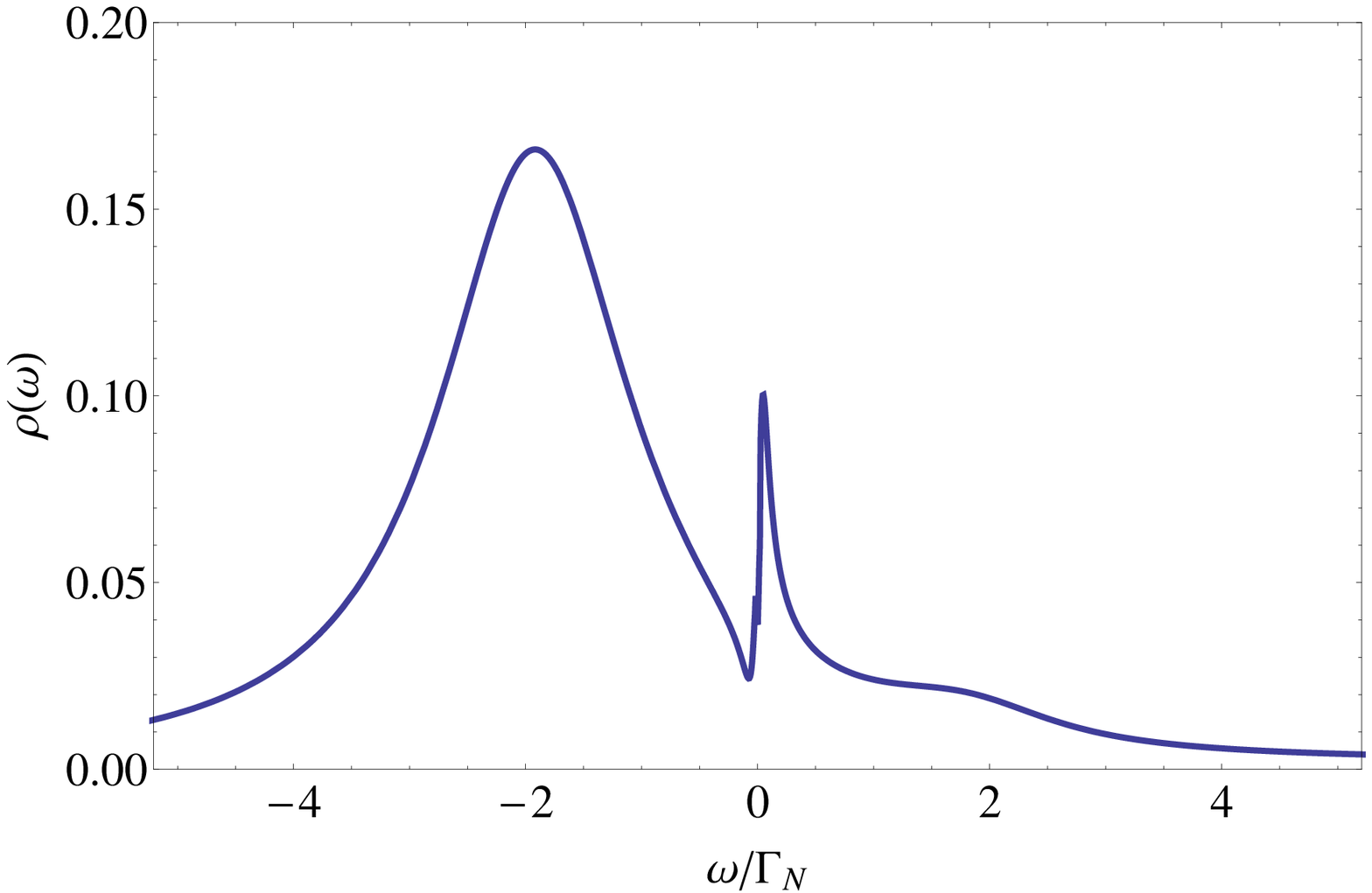}}
\caption{Spectral function $\rho(\omega)$ of the correlated QD obtained 
in the Kondo regime for $\uparrow$ electrons using $\epsilon=-2\Gamma_N$, 
$U=25\Gamma_N$,  $\Gamma_S=4\Gamma_N$, $t_{m}=0.3\Gamma_{N}$. The upper 
panel refers to $\epsilon_m=\Gamma_N$ and the bottom one to $\epsilon_m=0$.}
\label{Kondo_up}
\end{figure}

\begin{figure}
\epsfxsize=7cm\centerline{\epsffile{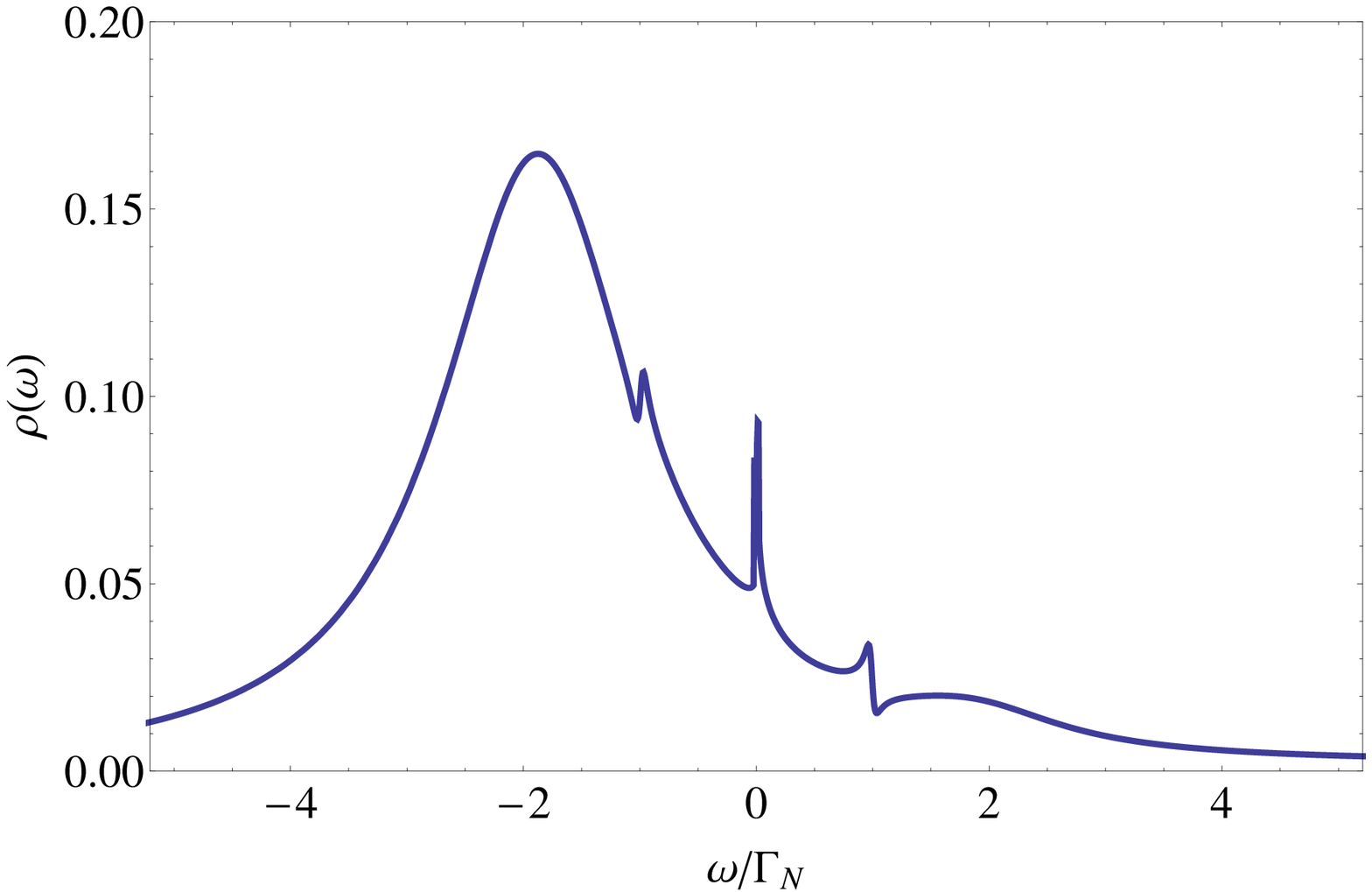}}
\epsfxsize=7cm\centerline{\epsffile{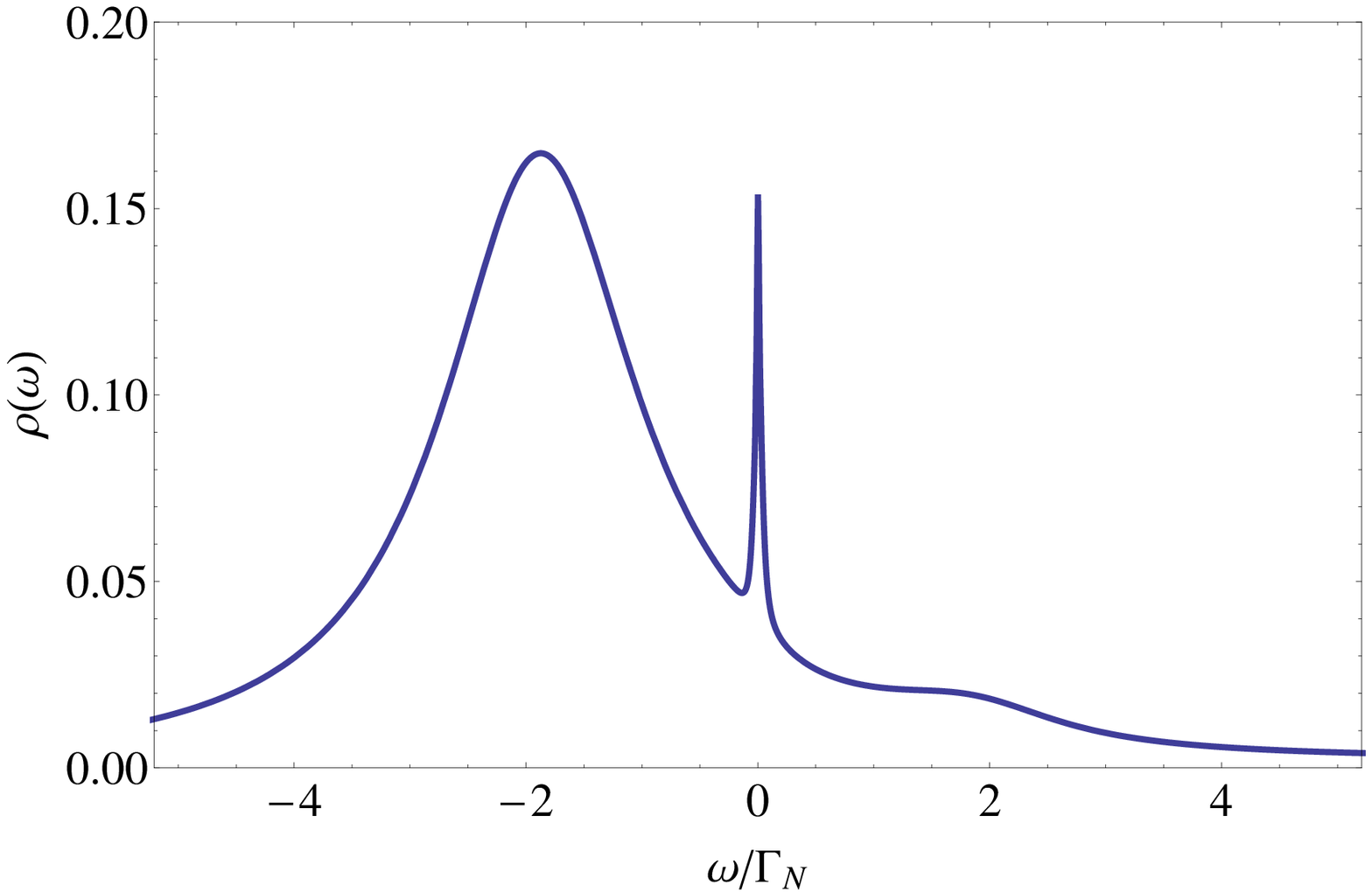}}
\vspace{0.0cm}
\caption{Spectral function of the spin $\downarrow$ electrons obtained 
for the same set of model parameters as in figure \ref{Kondo_up}.}
\label{Kondo_down}
\end{figure}

Influence of the side-coupled Majorana quasiparticle on the spin-resolved spectral 
functions $\rho_{\sigma}(\omega)$ of the correlated QD is illustrated in figures 
\ref{Kondo_up} and \ref{Kondo_down}. The upper panels correspond to the case of 
overlapping Majorana modes $\epsilon_{m}=\Gamma_{N}$. In analogy to the noninteracting 
situation (Fig.\ \ref{overlap}) we observe the fractional Fano and anti-Fano 
lineshapes appearing at $\omega=\pm \epsilon_m$ in the spectrum of spin $\uparrow$ 
and $\downarrow$ electrons, respectively. For $\downarrow$ electrons we also notice 
that both anti-Fano resonances are much less pronounced as compared to $U=0$ case. 
This is a consequence of the strong Coulomb interactions suppressing the on-dot 
pairing, that is indirectly responsible for the interferometric structures 
in the spectrum of $\downarrow$ electrons.

The most intriguing case occurs for $\epsilon_m=0$, when the Kondo and 
interferometric structures coincide at exactly the same energy. Spectrum 
of $\uparrow$ electrons, that are directly  coupled to the Majorana 
quasiparticle clearly show the dominant and destructive influence of 
the quantum interference on the Kondo state (see the bottom panel in 
Fig.\ \ref{Kondo_up}, where the Kondo peak is completely washed out). 
As regards the spectrum of $\downarrow$ electrons, the anti-Fano 
interferometric structure constructively combines with the Kondo 
peak, enhancing the zero-energy feature.

Scattering mechanism driven by the zero-energy Majorana quasiparticle 
side-attached to the correlated quantum dot has thus very interesting 
effect on the Kondo state. For the spin $\uparrow$ sector (directly 
coupled to the Majorana quasiparticle) the ongoing quantum interference 
has destructive character. In other words, the fractional Fano-type 
resonance induced by the side-coupled Majorana quasiparticle is robust 
against the Kondo peak. On contrary, in the spin $\downarrow$ sector 
(where electrons are not directly coupled to the Majorana quasiparticle) 
the Kondo state is promoted by the quantum interference. Such exotic 
spin-resolved quantum interference effects might be useful for experimental
detection of the Majorana quasiparticle.
From a physical point of view, this spin-resolved screening effects of the correlated
quantum dot is due to the following mechanism: the spin $\uparrow$ electrons 
`leak' into the side-coupled Majorana structure (hence there is less spin 
$\uparrow$ to be screened), whereas the on-dot pairing compensates such 
loss by enhancing the density of $\downarrow$ electrons  whose screening 
is effectively pronounced. 

\begin{figure}
\epsfxsize=8cm\centerline{\epsffile{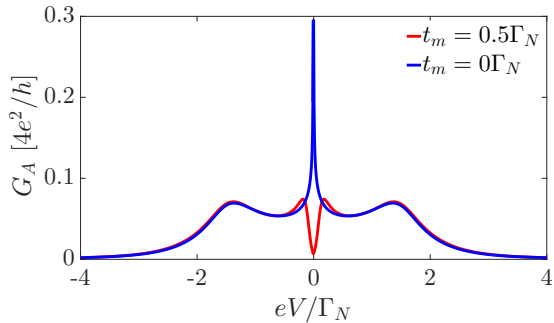}}
\caption{Differential conductance $G_{A}$ of the subgap current (\ref{I_A}) 
as a function of the applied voltage $V$ for $\epsilon=-1.5\Gamma_{N}$,
$U=25\Gamma_{N}$, $\Gamma_{S}=4\Gamma_{N}$ and $\epsilon_{m}=0$. The blue 
line corresponds to $t_{m}=0$ and the red curve is obtained for $t_{m}=
0.5\Gamma_{N}$.}
\label{new1}
\end{figure}

Practical observation of the Majorana and Kondo signatures would be feasible
only indirectly, by measuring a differential conductance $G_{A}=dI_{A}(V)/dV$ 
of the net subgap current (\ref{I_A}). Since both effects appear at zero energy, 
they shhuld be manifested in the linear conductance (i.e.\  at $V=0$).   
In absence of the side-attached Majorona quasiparticle (blue line in 
Fig.\ \ref{new1}) we indeed observe a zero-bias enhancement driven by the subgap 
Kondo effect, that has been reported experimentally by several groups 
\cite{Deacon-2010,Lohneysen-2012,Chang-2013,Aguado-2013}. For the case, when 
Majorana quasiparticle is side-coupled to the interracial QD (red line in 
Fig.\ \ref{new1}) there appears the dip at $V = 0$  instead of the previous 
enhancement. Such destructive effect could thus distinguish between 
the Kondo and Majorana features. Furthermore, in realistic situation  
the magnetic field (which is necessary for inducing the zero-energy mode 
of TS wire) would additionally split the Kondo resonance, shifting it 
from $V=0$. In the configuration discussed here the zero-bias dip 
caused by the Majorana quasiparticle is hence robust against eventual 
spectroscopic feature of the Kondo effect.

\section{Summary}

We studied interferometric structures induced by the Majorana 
quasiparticle side-coupled to the quantum dot on interface between 
the superconducting and normal electrodes. Due to the superconducting 
proximity effect such lineshapes appear simultaneously in both spin sectors, 
even though only one of the spins is directly coupled to the Majorana state. 
For each spin component, however, they are manifested differently. 

The subgap spectrum of $\uparrow$ electrons (directly coupled to the 
Majorana quasiparticle) is characterized by the fractional Fano-type 
lineshapes. Their fractionality is caused by the fact that Majorana 
quasiparticle is half of a true electronic state. On the other hand, 
the spectrum of opposite spin electrons is characterized by anti-Fano 
lineshapes appearing at the same energies as for $\uparrow$ electrons. 
Such quantum interference does effectively yield 
different (spin-resolved) Andreev transmittances.

We also confronted the spin-resolved interferometric features with 
the Kondo effect (caused by the strong correlations). We found that 
the side-coupled Majorana quasiparticle can either suppress or enhance 
the Kondo effect, depending on the spin orientation. Screening of 
electrons that are directly coupled to the Majorana quasiparticle
is practically destroyed by the quantum interference, whereas for 
the opposite spin component reveals substantial enhancement of the 
Kondo peak. We hope that our results can stimulate experimental 
efforts to verify the spin-selective influence of the Majorana 
quasiparticles on the Kondo state.

\section*{Acknowledgments}

We acknowledge P.\ Stefa\'nski for valuable discussions and
thank K.J.\ Kapcia for technical assistance.
This work is supported by the National Science Centre in Poland 
via project DEC-2014/13/B/ST3/04451 (TD).

\appendix

\section{Derivation of Green's functions}

In this appendix we outline procedure for determination of the matrix Green's 
function (\ref{Gr44}). Starting from the model Hamiltonian (\ref{HAnd}) we
consider the uncorrelated quantum dot $U=0$ coupled to between the metallic (N) 
and superconducting (S) electrodes and additionally side-coupled to the edge 
of topological wire.  In the deep subgap regime the superconducting 
electrode induces the static on-dot pairing and (\ref{HAnd}) 
simplifies to 
\begin{eqnarray}
H&=&
H_{N}+ H_{T,N} + \frac{\Gamma_S}{2} (d_{\uparrow}^{\dagger} 
d_{\downarrow}^{\dagger} + d_{\downarrow} d_{\uparrow})
+ \sum_{\sigma} \epsilon d_{\sigma}^{\dagger}d_{\sigma} 
\nonumber \\&+&
t_{m} (d_{\uparrow}^{\dagger}-d_{\uparrow})(f^{\dagger}+f)
+ \epsilon_{m}(f^{\dagger}f+1/2) .
\end{eqnarray}

Fourier transform of the retarded Green's function can be computed 
from the equation of motion  $\omega \langle
\langle A ; B \rangle\rangle=\langle [A , B ]_{+}\rangle + \langle\langle
[A,H]_{-} ; B\rangle\rangle$ where $\pm$ denote anticommutator/commutator, 
respectively. The particle propagator $\langle\langle d_{\uparrow} ; 
d_{\uparrow}^{\dagger} \rangle\rangle$ for spin $\uparrow$ electrons 
of the central QD is mixed with the operators of TS wire and with 
the anomalous Green's function 
\begin{eqnarray}
\left( \omega-\epsilon+i\frac{\Gamma_N}{2} \right) \langle\langle 
d_{\uparrow}; d_{\uparrow}^{\dagger} \rangle\rangle = 1 - 
\frac{\Gamma_S}{2} \langle\langle d_{\downarrow}^{\dagger}; 
d_{\uparrow}^{\dagger}\rangle\rangle
\nonumber \\ +
t_{m}\langle\langle \left( f^{\dagger} + f \right); 
d_{\uparrow}^{\dagger} \rangle\rangle 
\label{g1up}
\end{eqnarray}
As $\downarrow$ electrons are not directly coupled to TS wire the anomalous Green's 
function $\langle\langle d_{\downarrow}^{\dagger} ; d_{\uparrow}^{\dagger} 
\rangle\rangle$ does not generate any propagator containing $f$ operators. 
Using EOM, it can be expressed via the hole propagator
\begin{eqnarray}
\left( \omega+\epsilon+i\frac{\Gamma_N}{2} \right) \langle\langle 
d_{\downarrow}^{\dagger} ; d_{\uparrow}^{\dagger} \rangle\rangle 
=-\frac{\Gamma_S}{2} \langle\langle d_{\uparrow} ; d_{\uparrow}^{\dagger} 
\rangle\rangle
\label{A3}
\end{eqnarray}
Using (\ref{A3}) we can rewrite equation (\ref{g1up}) as   
\begin{eqnarray}
\left( \omega-\epsilon+i\frac{\Gamma_N}{2} -\frac{(\Gamma_S/2)^2}
{\omega+\epsilon+i\Gamma_N/2} \right)\langle\langle d_{\uparrow} 
; d_{\uparrow}^{\dagger} \rangle\rangle = \nonumber \\ 
1 + t_{m} \langle\langle f^{\dagger}; d_{\uparrow}^{\dagger} 
\rangle\rangle + t_{m} \langle\langle f; d_{\uparrow}^{\dagger} 
\rangle\rangle .
\label{g1up2}
\end{eqnarray}
We can notice that expression in the bracket on left hand side 
is the inverse particle propagator for N-QD-S system in absence 
of the TS wire. For brevity  we denote it by symbol $a$, that 
is presented in Eqn (\ref{a_symbol}).

The other Green's functions, where $d_{\sigma}^{\dagger}$ is mixed 
with $f$ and $f^{\dagger}$ operators can be found from the EOM as
\begin{eqnarray}
(\omega-\epsilon_m)\langle\langle f; d_{\uparrow}^{\dagger} \rangle\rangle
&=& t_{m} \langle\langle d_{\uparrow}; d_{\uparrow}^{\dagger} \rangle\rangle
- t_{m}\langle\langle d_{\uparrow}^{\dagger}; d_{\uparrow}^{\dagger} 
\rangle\rangle , 
\label{m} \\
(\omega+\epsilon_m)\langle\langle f^{\dagger}; d_{\uparrow}^{\dagger} \rangle\rangle
&=& t_{m} \langle\langle d_{\uparrow}; d_{\uparrow}^{\dagger} \rangle\rangle
-t_{m} \langle\langle d_{\uparrow}^{\dagger}; d_{\uparrow}^{\dagger} \rangle\rangle .
\label{n}
\end{eqnarray}
Equations (\ref{m},\ref{n}) generate the new anomalous function 
$\langle\langle d_{\uparrow}^{\dagger} d_{\uparrow}^{\dagger} \rangle\rangle$. 
We write down the equation of motion for this function
\begin{eqnarray}
\left( \omega+\epsilon+i\frac{\Gamma_N}{2}\right) \langle\langle 
d_{\uparrow}^{\dagger} ; d_{\uparrow}^{\dagger} \rangle\rangle 
=\frac{\Gamma_S}{2} \langle\langle d_{\downarrow} ; 
d_{\uparrow}^{\dagger}\rangle\rangle 
\nonumber \\
-t_{m} ( \langle\langle f^{\dagger}; d_{\uparrow}^{\dagger} \rangle\rangle 
+ \langle\langle f; d_{\uparrow}^{\dagger} \rangle\rangle)
\label{gwierd}
\end{eqnarray}
and determine the new function $\langle\langle d_{\downarrow} ; 
d_{\uparrow}^{\dagger}\rangle\rangle$  as
\begin{eqnarray}
\left( \omega-\epsilon+i\frac{\Gamma_N}{2} \right) \langle\langle 
d_{\downarrow} ; d_{\uparrow}^{\dagger} \rangle\rangle = 
\frac{\Gamma_S}{2} \langle\langle d_{\uparrow}^{\dagger} ;
d_{\uparrow}^{\dagger} \rangle\rangle .
\end{eqnarray}
Now the propagator $ \langle \langle d_{\uparrow}^{\dagger} 
;d_{\uparrow}^{\dagger} \rangle\rangle$ can be represented as
\begin{eqnarray}
\left( \omega+\epsilon+i\Gamma_N/2 -\frac{(\Gamma_S/2)^2}{\omega
-\epsilon+i\Gamma_N/2} \right) \langle\langle d_{\uparrow} ; 
d_{\uparrow}^{\dagger} \rangle\rangle =
\nonumber \\ 
-t_{m}\left( \langle\langle f^{\dagger}; d_{\uparrow}^{\dagger} 
\rangle\rangle + \langle\langle f; d_{\uparrow}^{\dagger} 
\rangle\rangle \right) .
\label{gwierd2}
\end{eqnarray}
Expression appearing in a bracket on the left hand side is the inverse 
hole propagator of N-QD-S system without TS wire. We have denoted 
it by symbol $b$ in the main text, that is explicitly given 
by Eqn (\ref{b_symbol}).

Finally, we obtain the following set of equations 
\begin{eqnarray}
a \langle\langle d_{\uparrow} ; d_{\uparrow}^{\dagger} \rangle\rangle
&=&1+t_{m}  \langle\langle f^{\dagger}; d_{\uparrow}^{\dagger} \rangle\rangle 
+ t_{m} \langle\langle f; d_{\uparrow}^{\dagger} \rangle\rangle ,
\nonumber \\
b \langle\langle d_{\uparrow}^{\dagger} ; d_{\uparrow}^{\dagger} \rangle\rangle
&=&-t_{m}  \langle\langle f^{\dagger}; d_{\uparrow}^{\dagger} \rangle\rangle 
-t_{m} \langle\langle f; d_{\uparrow}^{\dagger} \rangle\rangle ,
\nonumber \\
(\omega-\epsilon_m)\langle\langle f; d_{\uparrow}^{\dagger} \rangle\rangle
&=& t_{m}  \langle\langle d_{\uparrow}; d_{\uparrow}^{\dagger} \rangle\rangle
- t_{m} \langle\langle d_{\uparrow}^{\dagger}; d_{\uparrow}^{\dagger} \rangle\rangle ,
\nonumber \\
(\omega+\epsilon_m)\langle\langle f^{\dagger}; d_{\uparrow}^{\dagger} \rangle\rangle
&=& t_{m} \langle\langle d_{\uparrow}; d_{\uparrow}^{\dagger} \rangle\rangle-t_{m}
\langle\langle d_{\uparrow}^{\dagger}; d_{\uparrow}^{\dagger} \rangle\rangle  .
\nonumber
\end{eqnarray}
Using the  abbreviations $m\equiv (\omega-\epsilon_m)$, $n\equiv (\omega+\epsilon_m)$ 
and denoting $W \equiv abmn-2t_{m}^2\omega(a+b)$ these Green's functions can be recast 
in the following matrix form
\begin{eqnarray}
\left( \begin{array}{cc}  
\langle\langle d_{\uparrow}; d_{\uparrow}^{\dagger} \rangle\rangle  & 
\langle\langle f; d_{\uparrow}^{\dagger} \rangle\rangle \\
\langle\langle f^{\dagger}; d_{\uparrow}^{\dagger} \rangle\rangle & 
\langle\langle d_{\uparrow}^{\dagger}; d_{\uparrow}^{\dagger} \rangle\rangle
 \end{array}\right)=\frac{1}{W}
\left( \begin{array}{cc}  
bmn-2t_{m}^2\omega  & bnt_{m} \\
bmt_{m} & -2t_{m}^2 \omega
\end{array}\right) 
\nonumber \\
\end{eqnarray}
The entire matrix presented in Eqn (\ref{Gr44}) can be obtained 
by solving 4 similar sets of the equations.

\section{Spin-dependent coupling to normal QD}

\begin{figure}
\epsfxsize=8cm\centerline{\epsffile{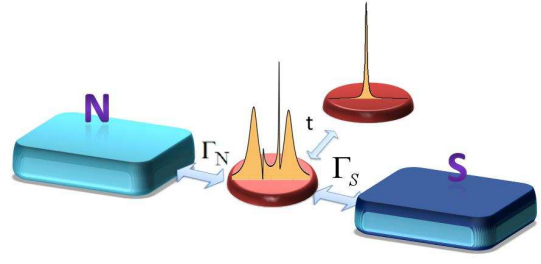}}
\caption{Schematic view of the Fano and anti-Fano interference patterns 
induced in the T-shape setup with both normal quantum dots for 
$t_{\uparrow}=t_{\downarrow}$.} 
\label{scheme_DQD}
\end{figure}

To distinguish the consequences caused by the fact that tunneling to 
the MQD involves only the spin $\uparrow$ electrons from other effects 
due to their specific  Majorana-type nature we examine here the setup 
in which TS is replaced by the usual quantum dot (QD$_{2}$) 
\begin{eqnarray}
H_{MQD}\rightarrow H_{QD_2}= \sum_{\sigma} \epsilon_2 d^{\dagger}_{2 \sigma} 
d_{2 \sigma} + \sum_{\sigma} t_{\sigma}(d^{\dagger}_{\sigma} d_{2 \sigma}+h.c.) .
\nonumber \\
\label{toy_model_bis}
\end{eqnarray}
Our previous study \cite{Baranski-2011} of such normal double quantum dot 
(DQD) in the T-shape configuration  has been done for the spin independent 
couplings $t_{\uparrow}=t_{\downarrow}$. Under such circumstances we have 
obtained the Fano and anti-Fano resonances, schematically displayed in 
Fig.\ \ref{scheme_DQD}.

\begin{figure}
\epsfxsize=7cm\centerline{\epsffile{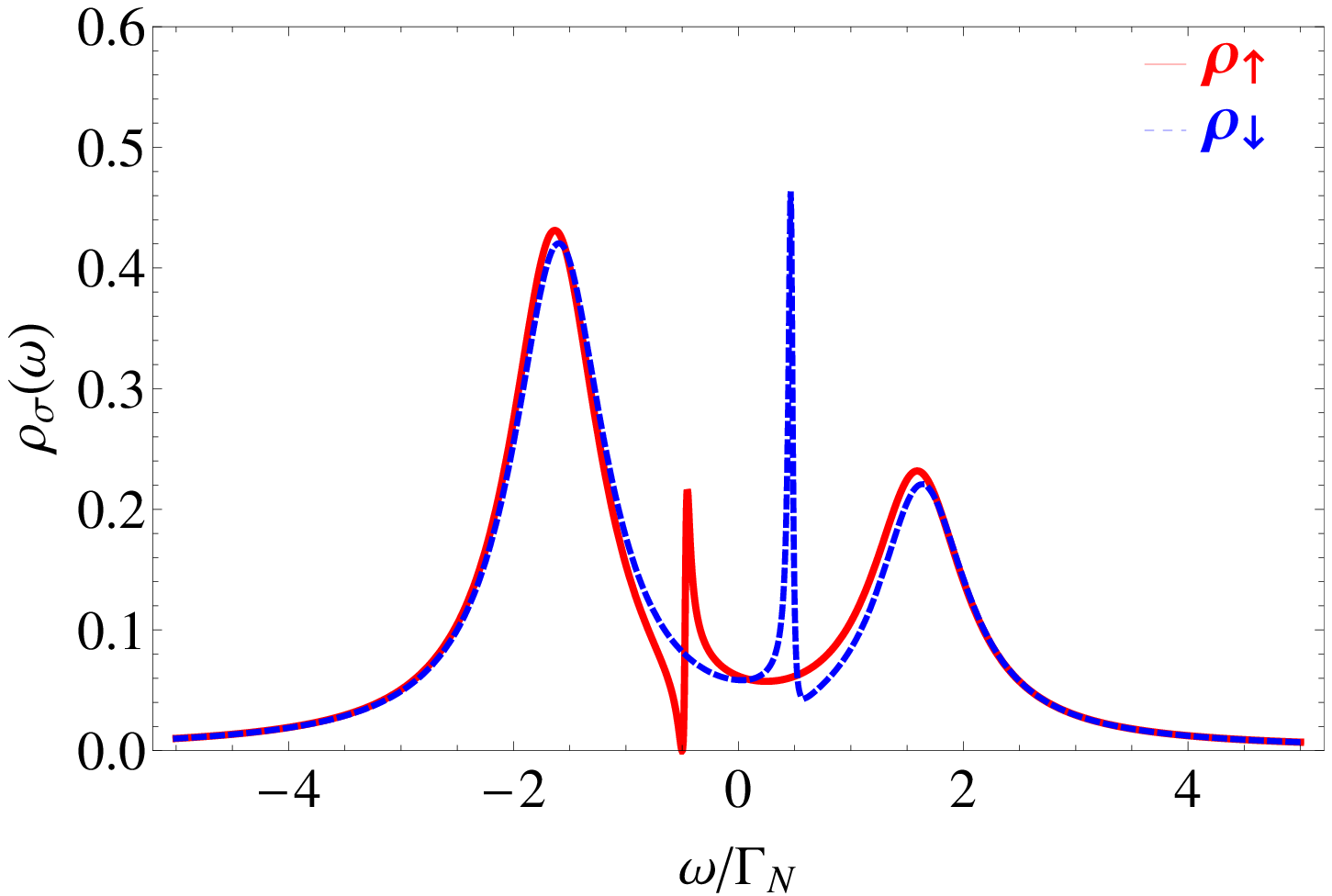}}
\epsfxsize=7cm\centerline{\epsffile{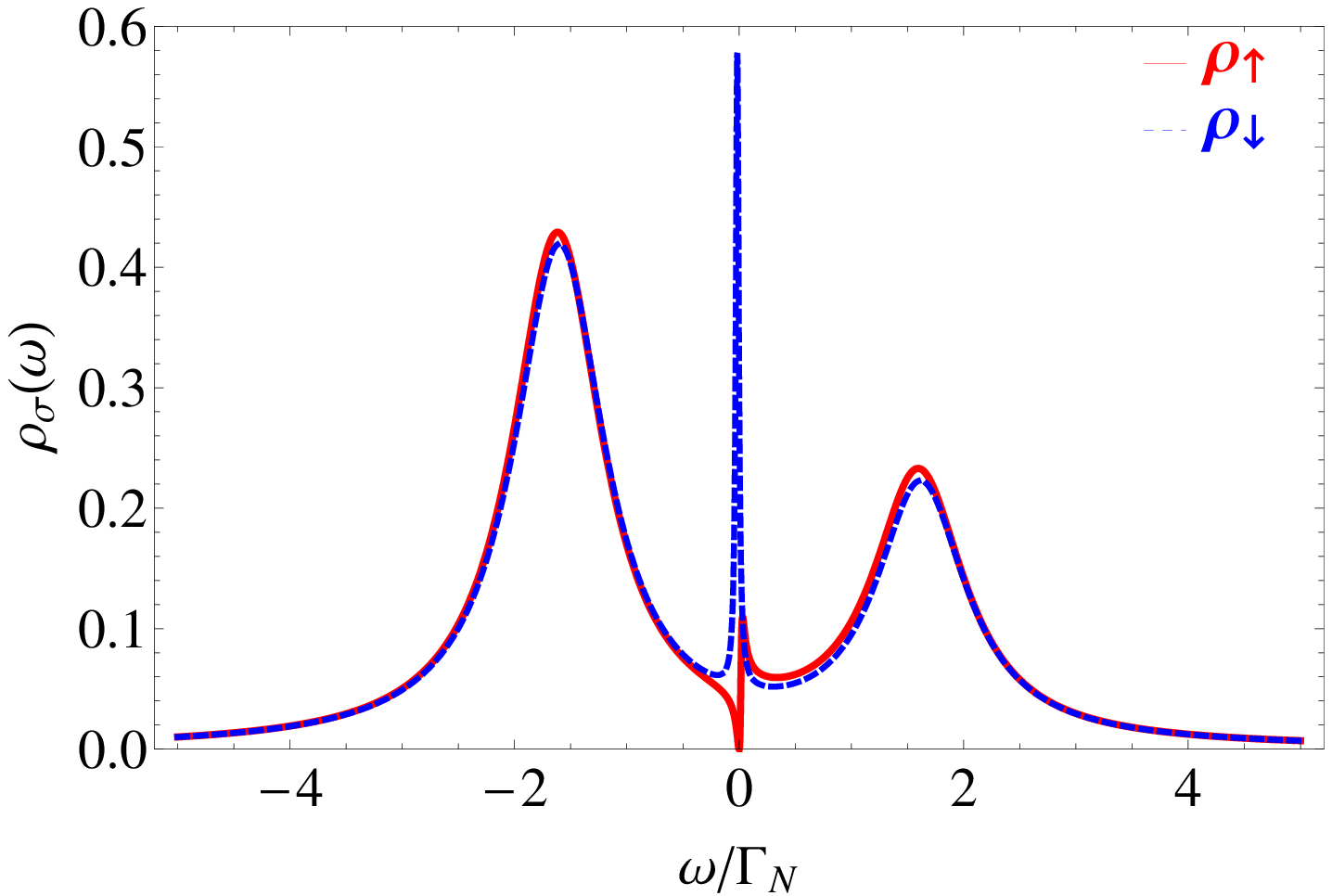}}
\caption{Spectral function $\rho_{\sigma}(\omega)$
of the uncorrelated QD asymmetrically coupled  ($t_{\uparrow}=0.3 
\Gamma_N$, $t_{\downarrow}=0$) to the normal QD$_{2}$. 
The solid (red ) line refers to $\uparrow$ electrons and the dashed (blue)  
once to spin $\downarrow$ electrons. The results are obtained 
for the model parameters $\Gamma_S=3\Gamma_N$, $\epsilon=-0.5\Gamma_N$ 
and $\epsilon_2=0$. We notice the usual Fano-type pattern for $\uparrow$ 
electrons (directly coupled to QD) accompanied by the anti-Fano feedback 
for spin $\downarrow$ electrons (due to the on-dot pairing).
Top panel refers to $\epsilon_2=-0.5\Gamma_N$ and the bottom one to
$\epsilon_2=0$.}
\label{ordsep}
\end{figure}

In this Appendix we address the spin-polarized coupling $t_{\uparrow} 
\neq t_{\downarrow}$, focusing on the limit of vanishing $t_{\downarrow}$. 
Fourier transform of the retarded Green's function for the uncorrelated 
central quantum dot is expressed by \cite{Baranski-2011} 
\begin{eqnarray} 
&&\left[ \begin{array}{cc}  
\langle\langle d_{\sigma} ; d_{\sigma}^{\dagger} \rangle\rangle  & 
\langle\langle d_{\sigma} ; d_{\bar{\sigma}} \rangle\rangle   \\
\langle\langle d_{\bar{\sigma}}^{\dagger} ; d_{\sigma}^{\dagger} \rangle\rangle   &  
\langle\langle d_{\bar{\sigma}}^{\dagger} ; d_{\bar{\sigma}} \rangle\rangle  
\end{array}\right] 
\label{ORD_G} \\
&=& \left( \begin{array}{cc}  
\omega-\epsilon +\frac{i\Gamma_N}{2} - \frac{t^{2}_{\sigma}}{\omega-\epsilon_2}  
&  \frac{\Gamma_S}{2} \\ \frac{\Gamma_S}{2} &  
\omega+\epsilon + \frac{i\Gamma_N}{2} -\frac{t^{2}_{\bar{\sigma}}}{\omega+\epsilon_2}
\end{array}\right)^{-1}
\nonumber
\end{eqnarray}  
where $\bar{\sigma}$ is inverse spin to $\sigma$. For the weak identical couplings 
$t_{\uparrow}=t_{\downarrow}$  the spectral function of central quantum dot 
$\rho_{\sigma}(\omega) = -\pi^{-1} \mbox{\rm Im} \langle\langle d_{\sigma} 
; d_{\sigma}^{\dagger} \rangle\rangle$ is characterized by two interferometric 
structures at the QD$_{2}$ level  and on the opposite side of a Fermi level 
\cite{Baranski-2011}. The feature at $\omega=\epsilon_2$ has the usual Fano-type 
resonant lineshape \cite{Zitko-2010}, whereas its companion at $-\epsilon_{2}$ 
has anti-resonant (antiFano) shape. Obviously for $t_{\uparrow}=t_{\downarrow}$  
the spectral functions $\rho_{\sigma}(\omega)$ of both spins are identical.

When the tunneling of $\downarrow$ electrons is forbidden ($t_{\downarrow}=0$), 
the toy model (\ref{toy_model_bis}) is closely  analogous to the original setup 
(Fig.\ \ref{schematics}) with only spin $\uparrow$ of the central quantum dot 
coupled to TS wire. In such case there survives the single interferometric 
structure in each of the spectral functions $\rho_{\sigma}(\omega)$. For 
$\uparrow$ electrons (directly coupled to the side-attached dot) we observe 
the Fano-type interference pattern at $\varepsilon_{2}$ and for the opposite 
spin $\downarrow$ electrons there appears the anti-Fano structure at 
$-\varepsilon_{2}$. This result can be understood if we anticipate that 
the anti-Fano feature is an indirect response of the spin $\uparrow$ 
electrons. In other words, even though the spin $\downarrow$ electrons are not 
directly coupled to the side-attached quantum dot, due to the induced local 
pairing they `feel' a feedback from the opposite ($\uparrow$) spin  electrons.

\begin{figure}
\epsfxsize=7cm\centerline{\epsffile{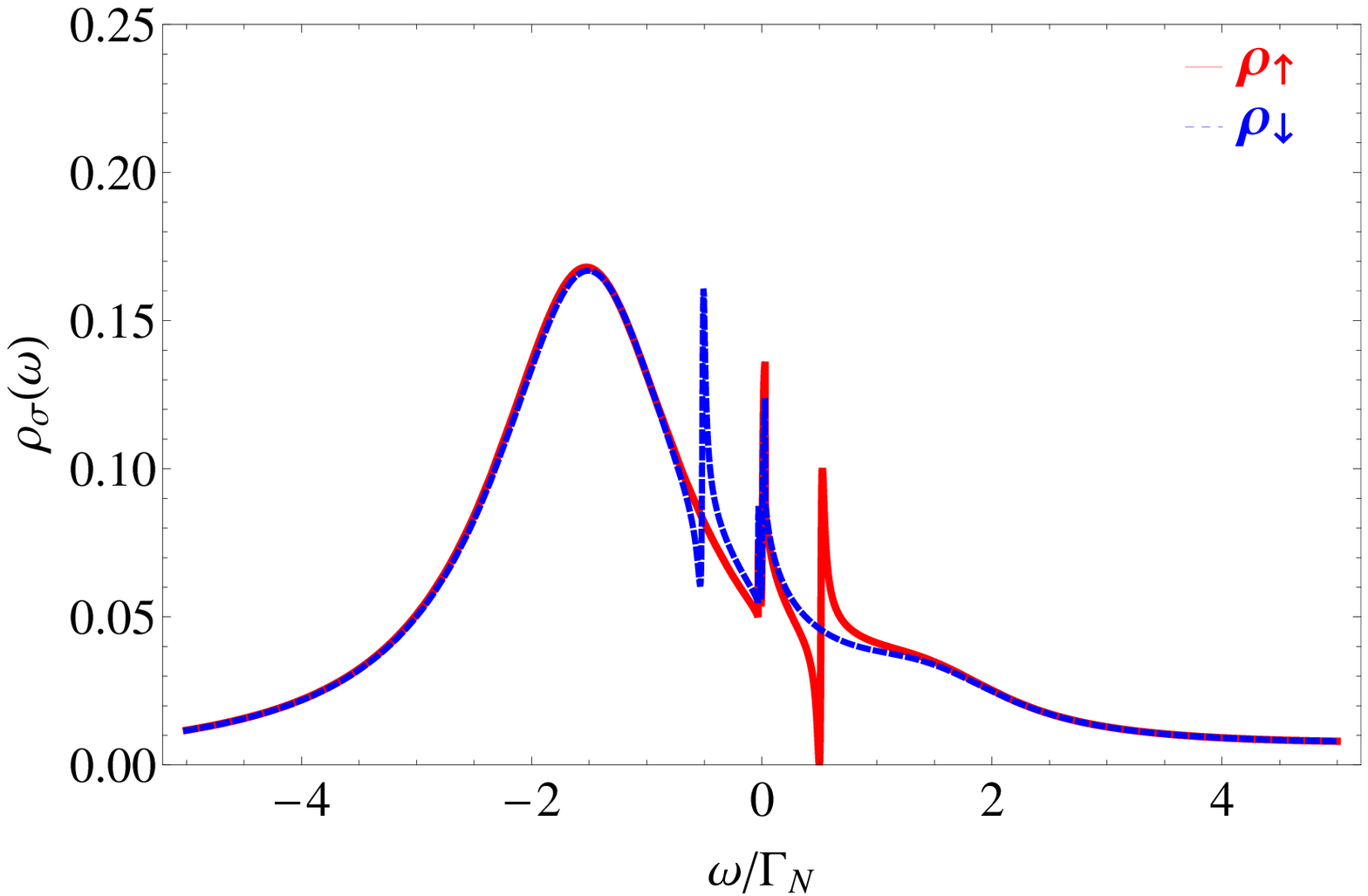}}
\epsfxsize=7cm\centerline{\epsffile{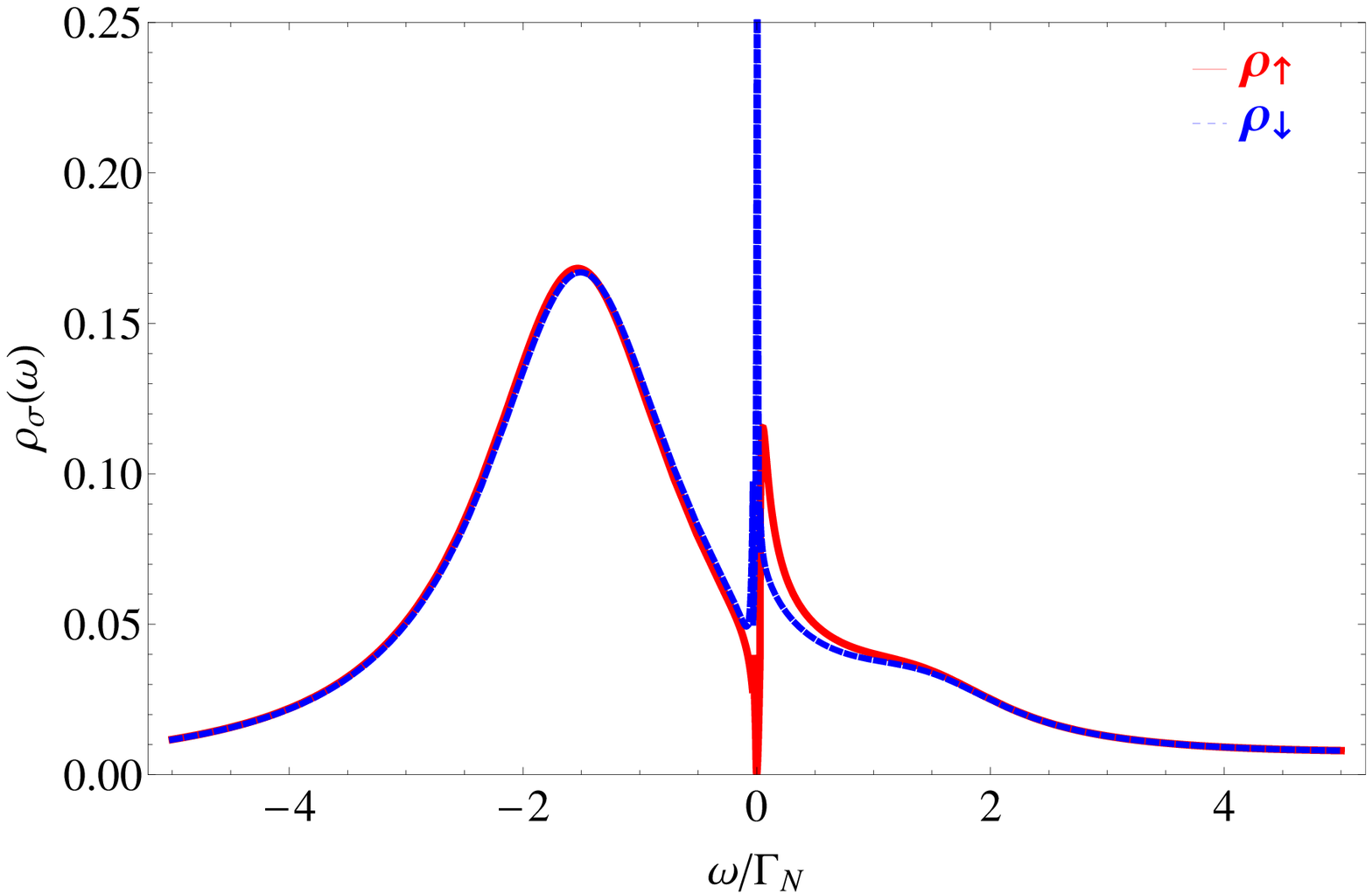}}
\caption{Spectral function the correlated QD side-coupled to the normal QD$_{2}$ 
obtained for  $\Gamma_S=4\Gamma_N$, $U=15\Gamma_N$, 
$k_{B}T=0.005\Gamma_N$. The solid (blue) line refers to $\uparrow$ and 
the dashed (red) line to $\downarrow$ electrons. Top panel refers to
$\epsilon_2=0.5\Gamma_N$ and the bottom one to $\epsilon_2=0$.}
\label{Kondo_Ord1}
\end{figure}

The upper panel in figure \ref{ordsep} illustrates the spin-resolved 
spectral functions of the uncorrelated central quantum dot asymmetrically 
coupled to the normal QD$_{2}$ whose energy level is $\epsilon_2 \neq 0$, 
where interferometric features appear either in the particle or hole regions.
The bottom panel in figure \ref{ordsep} corresponds to  the situation with 
$\epsilon_2=0$. In this case the Fano and anti-Fano lineshapes appear at 
the same position what is partly similar to the case with Majorana quasiparticle.

Figure \ref{Kondo_Ord1} shows the spectral function $\rho_{\sigma}(\omega)$ 
obtained fairly below the Kondo temperature $T_{K}$ for spin $\uparrow$ 
(red line) and $\downarrow$ (blue line) electrons. The approximation 
described in Sec.\ III.A cannot reliably reproduce the low energy structure 
of the Kondo peak $|\omega|\geq k_{B}T_{K}$, therefore our results should 
be treated only qualitatively.

In the case when $\epsilon_{2}$ is far from the Kondo
peak (top panel) we observe that the Fano-type resonance 
(seen in $\rho_{\uparrow}(\omega)$ at $\epsilon_{2}$) and its anti-Fano
companion (present in $\rho_{\downarrow}(\omega)$ at $-\epsilon_{2}$)
practically coexist with the zero-energy Kondo peak.
The situation changes dramatically, when  energy $\epsilon_{2}$ 
coincides with the Fermi level (bottom panel). In both spin 
sectors the Kondo peak is then completely destroyed by the interferometric 
lineshape. This effect proves that the quantum interference is dominant, 
whenever it coincides with the  Kondo peak. Let us notice that 
such tendency is distinct from the interferometric features induced by 
the Majorana quasiparticle (Figs \ref{Kondo_up} and \ref{Kondo_down}).

\section{Influence of the trivially paired dot}

\begin{figure}
\epsfxsize=6cm\centerline{\epsffile{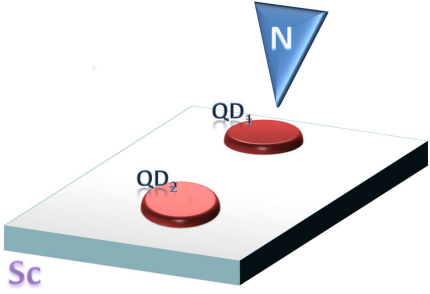}}
\caption{Schematic view of STM-type configuration, where transport can 
occur between the normal tip (N) and the superconducting (Sc) substrate 
via the central quantum dot (QD$_{1}$) which is laterally coupled to another 
quantum dot (QD$_{2}$). Both dots absorb the $s$-wave (trivial) pairing.}
\label{trivial_QD}
\end{figure}

In analogy to Appendix B, we consider now the side-attached quantum 
dot QD$_{2}$ characterized by the usual $s$-wave pairing. Physically 
such situation can be achieved in STM-type configuration illustrated 
by Fig.\ \ref{trivial_QD}. The charge transport would occur between 
the normal tip (N) and the superconducting lead (Sc) through the 
quantum dot (QD$_{1}$) deposided on superconductor and laterally 
coupled to another quantum dot (QD$_{2}$), that rests on the same 
superconducting substrate. This setup can be described by the model 
similar to (\ref{toy_model_bis}) with the $s$-wave pairing imposed 
on QD$_{2}$. Formally we use the following model
$H_{QD_2}=\sum_{\sigma} \epsilon_2 d^{\dagger}_{2 \sigma} 
d_{2 \sigma} - \left( \frac{\Gamma_{S2}}{2} d^{\dagger}_{2 \uparrow}
d^{\dagger}_{2 \downarrow} + \mbox{\rm h.c.} \right) 
+ \sum_{\sigma} \left( t d^{\dagger}_{1 \sigma} d_{2 \sigma}
+ \mbox{\rm h.c.} \right)$, where $\Gamma_{S2}$ describes 
the effective coupling of QD$_{2}$ to the $s$-wave superconducting
reservoir.

\begin{figure}
\epsfxsize=7cm\centerline{\epsffile{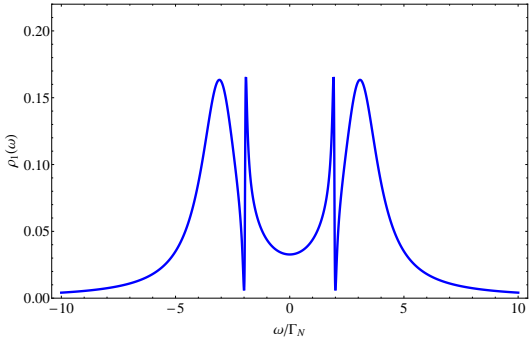}}
\caption{Spectral function $\rho_{1}(\omega)$ of the interfacial quantum 
dot obtained for $\epsilon=0$, $\Gamma_{S}=6\Gamma_{N}$, $\epsilon_{2}=0$
$\Gamma_{S2}=4\Gamma_{N}$, $t=0.3\Gamma_{N}$, neglecting the correlations.}
\label{transport_trivial}
\end{figure}

Key difference between the TS wire (that hosts the Majorana mode) and 
the QD$_{2}$ (that absorbs the usual $s$-wave superconductivity) can 
be observed in the electronic spectra. Majorana quasiparticle emerges 
at $\omega=0$, whereas the fermionic Shiba/Andreev states of QD$_{2}$ 
are formed at finite energies $\omega=\pm \sqrt{\epsilon_{2}^{2}+
(\Gamma_{S2}/2)^2}$.Influence of QD$_{2}$ on the subgap spectrum  of 
the central QD$_{1}$ is visible away from the Fermi level. Figure 
\ref{transport_trivial} presents the spectral function $\rho_{1}
(\omega)$ of QD$_{1}$. As a matter of fact, its electronic spectrum 
qualitatively differs from the unique features due to Majorana 
quasiparticle, discussed in main part of this work (see Fig.\ 
\ref{Mdqd}). 

\begin{thebibliography}{11}
%

\bibitem{Volovik-1999}
   Volovik G 1999
   {\rm Fermion zero modes on vortices in chiral superconductors} 
   {\it JETP Lett.} {\bf 70} 609 
 
\bibitem{Read-2000}
   Read N and Green D 2000
   {\rm Paired states of fermions in two dimensions with breaking of parity and time-reversal symmetries and the fractional quantum Hall effect} 
   {\it Phys.\ Rev.\ B} {\bf 61} 10267 

\bibitem{Kitaev-2001}
    Kitaev A Y 2001  
   {\rm Unpaired majorana fermions in quantum wires}  
   {\it Phys.\ Usp.} {\bf 44} 131 


\bibitem{Alicea-12}
Alicea J  2012
{\rm New directions in the pursuit of Majorana fermions in solid state systems} 
{\it Rep. Prog. Phys.} {\bf 75} 076501 

\bibitem{Flensberg-12}
Leijnse M and Flensberg K 2012
{\rm Introduction to topological superconductivity and  Majorana fermions} 
{\it Semicond. Sci. Technol.} {\bf 27} 124003 

\bibitem{Stanescu-13}
Stanescu T D and  Tewari S 2013
{\rm Majorana fermions in semiconductor nanowires: fundamentals, 
modeling, and experiment} 
{\it J. Phys.: Condens. Matter} {\bf 25} 233201 

\bibitem{Beenakker-13}
Beenakker C W J 2013  
{\rm Search for majorana fermions in superconductors} 
{\it Annu. Rev. Condens. Matt. Phys.} {\bf 4} 113 

\bibitem{Franz-15}
Elliot S R and Franz M 2015 
{\rm Colloquium: Majorana fermions in nuclear, particle, and solid-state physics} 
{\it Rev. Mod. Phys} {\bf 87} 137 

\bibitem{DasSarma-2016}
Liu X, Li X, Deng D -L, Liu X -J  and Das Sarma  S 2016
{\rm Majorna spintronics} 
{\it Phys.\ Rev.\ B} {\bf 94} 014511 


\bibitem{Tewari-2007}
Tewari S, Das Sarma S, Nayak C, Zhang C W and Zoller P 2007
{\rm Quantum computation using vortices and Majorana zero modes
of a $p_{x}+ip_{y}$ superfluid of fermionic cold atoms} 
{\it Phys.\ Rev.\ Lett.} {\bf 98} 010506 

\bibitem{Fu-2008}
Fu L and Kane C L 2008 {\rm Superconducting proximity effect
and Majorana fermions at the surface of a topological insulator}  
{\it Phys.\ Rev.\ Lett.} {\bf 100} 096407 

\bibitem{Nilsson-2008}
Nilsson J, Akhmerov A R and Beenakker C W J 2008
{\rm Splitting of a Cooper pair by a pair of Majorana bound state} 
{\it Phys.\ Rev.\ Lett.} {\bf 101} 120403 

\bibitem{Sato-2009}
 Sato M and  Fujimoto S 2009
{\rm Topological phases of noncentrosymmetric superconductors: edge states, 
Majorana fermions, and non-Abelian statistics} 
{\it Phys.\ Rev.\ B} {\bf 79} 094504 

\bibitem{Tworzydlo-2010}
Wimmer M, Akhmerov A R, Medvedyeva M V, Tworzyd\l o J and Beenakker C W J 2010
{\rm Majorana bound states without vortices in topological superconductors 
with electrostatic defects} 
{\it Phys.\ Rev.\ Lett.} {\bf 105} 046803  

\bibitem{Sau-2010}
Sau J D, Lutchyn R M, Tewari S and Das Sarma S 2010
{\rm Generic new platform for topological quantum computation using
semiconductor heterostructures} 
{\it Phys.\ Rev.\ Lett.} {\bf 104} 040502 

\bibitem{Oreg-2010}
Oreg Y, Refael G  and von Oppen F 2010 
{\rm Helical liquids and Majorana bound states in quantum wires}  
{\it Phys.\ Rev.\ Lett.} {\bf 105} 177002 

\bibitem{Lutchyn-2010} 
Lutchyn R M, Sau J D and Das Sarma S 2010 
{\rm Majorana fermions and a topological phase transition
in semiconductor-superconductor heterostructures}  
{\it Phys.\ Rev.\ Lett.} {\bf 105} 077001 

\bibitem{Choy-2011} 
Choy T P, Edge J M, Akhmerov A R and Beenakker C W J 2011 
{\rm Majorana fermions emerging from magnetic nanoparticles 
on a superconductor without spin-orbit coupling}  
{\it Phys.\ Rev.\ B} {\bf 84} 195442 


\bibitem{Aguado-2012}
San-Jose P, Prada E  and  Aguado R 2012
{\rm ac Josephson effect in finite-length nanowire junctions with Majorana modes} 
{\it Phys.\ Rev.\ Lett.} {\bf 108} 257001  


\bibitem{ultracold}
Jiang L, Kitagawa T, Alicea J, Akhmerov A R, Pekker D, 
Refael G, Cirac J I, Demler E, Lukin M D  and Zoller P 2011 
{\rm Majorana fermions in equilibrium and in driven cold-atom quantum wires} 
{\it Phys.\ Rev.\ Lett.} {\bf 106} 220402 


\bibitem{Mourik-12}
Mourik V, Zuo K, Frolov S M, Plissard S R, Bakkers   E P A M  
and Kouwenhoven  L P 2012
{\rm Signatures of Majorana fermions in hybrid superconductor-semiconductor 
nanowire devices} 
{\it Science} {\bf 336} 1003 

\bibitem{Yazdani-14}
Nadj-Perge S, Drozdov I K, Li  J, Chen  H,  Jeon S,
Seo J,  MacDonald  A H,  Andrei Bernevig B and Yazdani A 2014  
{\rm Observation of Majorana fermions in ferromagnetic atomic 
chains on a superconductor} 
{\it Science} {\bf 346} 602 

\bibitem{Kisiel-15}
Pawlak R, Kisiel M, Klinovaja J, Maier T, Kawai S, Glatzel T,   
Loss D and Meyer E 2015
{\rm Probing atomic structure and Majorana wave-functions in mono-atomic Fe-chains
on superconducting Pb-surface}
(arXiv:1505.06078)

\bibitem{Franke-15}
Ruby M, Pientka F, Peng Y, von Oppen F, Heinrich B W and Franke K J 2015 
{\rm End states and subgap structure in proximity-coupled chains of magnetic adatoms}
{\it Phys. Rev. Lett.} {\bf 115} 197204  


\bibitem{Liu-2012}
Liu J, Potter A C, Law K T  and Lee P A 2012
{\rm Zero-bias in the tunneling conductance of spin-orbit-coupling superconducting
wires with and without Majorana end-states}
{\it Phys. Rev. Lett.} {\bf 109} 267002  

\bibitem{Zitko-2015}
\v{Z}itko R, Lim J S, L\'opez R and Aguado R 2015
{\rm Shiba states and zero-bias anomalies in the hybrid normal-superconductor 
Anderson model} 
{\it Phys.\ Rev.\ B} {\bf 91} 045441 

\bibitem{Domanski-2016}
Doma\'nski T,  Weymann I, Bara\'nska M and G\'orski G 2016
{\rm Constructive influence of the induced electron pairing on the Kondo state} 
{\it Sci.\ Rep.} {\bf 6} 23336 

\bibitem{Rainis-2013}
Rainis D, Trifunovic L, Klinovaja J and Loss D 2013
{\rm Towards a realistic transport modelling in a superconducting nanowire
with Majorana fermions} 
{\it Phys.\ Rev.\ B} {\bf 87} 024515   


\bibitem{Chen-2014}
Chen H J and Zhu K D 2014
{\rm Nonlinear optomechanical detection for Majorana fermions via
a hybrid nanomechanical system} 
{\it Nanoscale Res.\ Lett.} {\bf 9} 166 

\bibitem{Lutchyn-2015}
Liu D E, Cheng M and Lutchyn R 2015
{\rm Probing Majorana physics in quantum-dot shot-noise experiments}
{\it Phys.\ Rev.\ B} {\bf 91} 081405(R) 

\bibitem{Flensberg-2016}
Hansen E B, Danon J and Flensberg K 2016
{\rm Phase-tunable Majorana bound states in a topological N-SNS junctions} 
{\it Phys.\ Rev.\ B} {\bf 93} 094501 

\bibitem{Baranger-2011}
Liu D E and Baranger H U 2011
{\rm Detecting a Majorana-fermion zero mode using a quantum dot} 
{\it Phys.\ Rev.\ B} {\bf 84} 201308(R) 

\bibitem{Leijnse-2014}
Leijnse M and Flensberg K 2014
{\rm Thermoelectric signatures of a Majorana bound state coupled to
a quantujm dot}
{\it New J.\ Phys.} {\bf 16} 015029 

\bibitem{Seridonio-2014}
Seridonio A C, Siqueira E C, Desotti F A, Mchado R S and Yoshida M 2014
{\rm Fano interference and a slight fluctuation of the Majorana hallmark} 
{\it J.\ Appl.\ Phys.} {\bf 115} 063706 

\bibitem{GongZhang-2014}
Gong W -J, Zhang S -F, Li Z -C, Yi G and Zheng Y -S 2014
{\rm Detection of a Majorana fermion zero mode by a T-shaped quantum-dot structure} 
{\it Phys.\ Rev.\ B} {\bf 89} 245413 

\bibitem{Jiang-2014}
Jiang C, Lu G  and Gong W -J 2014 
{\rm Tunable transport through a quantum dot chain with side-coupled
Majorana bound states} {\it J.\ Appl.\ Phys.} {\bf 116} 103704 

\bibitem{Desotti-2014}
Desotti F A, Ricco L S, de Souza M, Souza F M  and Seridonio A C 2014
{\rm Probing the antisymmetric Fano interference assisted by a Majorana fermion} 
{\it J.\ Appl.\ Phys.} {\bf 116} 173701 

\bibitem{Stefanski-2015}
Stefa\'nski P 2015
{\rm Signatures of Majorana states in electron transport through a quantum dot
coupled to topological wire} 
{\it Acta Phys.\ Polon.\ A} {\bf 127} 198 

\bibitem{Li-2015}
Li Z -Z, Lam C -H  and You J Q 2015
{\rm Probing Majorana bound states via counting statistics 
of a single electron transistor} 
{\it Sci.\ Rep.} {\bf 5} 11416 

\bibitem{Chirla-2016}
Chirla R and Moca C P 2016
{\rm Fingerprints of Majorana fermions in spin-resolved subgap spectroscopy} 
{\it Phys.\ Rev.\ B} {\bf 94} 045405 

\bibitem{Gong-2014}
Gong W -J, Zhang S -F, Li Z -C, Yi G and  Zheng Y -S 2014 
{\rm Andreev reflection in a T-shaped double-quantum-dot structure 
induced by Majorana bound states}
{\it J.\ Phys.\ Soc.\ Jpn.} {\bf 83} 034706

\bibitem{Wang-2016}
Wang S -X, Li Y -X,  Wang N and Liu J -J 2016
{\rm Andreev reflection in a T-shaped double quantum-dot with coupled 
Majorana bound states} 
{\it Acta Phys.\ Sin.}  {\bf 65} 137302  


\bibitem{He-2014}
He J J, Ng T K and Law K T 2014
{\rm Selective equal-spin Andreev reflections induced by Majorana fermions} 
{\it Phys.\ Rev.\ Lett.} {\bf 112} 037001 

\bibitem{Hu-2016}
Hu L -H,  Li C,  Xu D -H, Zhou Y and Zhang F -C 2016 
{\rm Theory for spin selective Andreev reflection in vortex core of topological
superconductor: Majorana zero modes on spherical surface and application to spin
polarized scanning tunneling microscope probe}
(arXiv:1607.03449)

\bibitem{Sun-2016}
Sun H -H et al 2016
{\rm Majorana zero modes detected with spin selective Andreev reflection 
in the vortex of a topological superconductor}
{\it Phys.\ Rev.\ Lett.} {\bf 116} 257003 


\bibitem{Alicea-16}
Mishmash R V,  Aasen D, Higginbotham A P and Alicea J 2016 
{\rm Approaching a topological phase transition in Majorana nanowires} 
{\it Phys.\ Rev.\ B} {\bf 93} 245404 


\bibitem{Baranski-2011}
Bara\'nski J and  Doma\'nski T 2011
{\rm Fano-type interference in quantum dots coupled between metallic 
and superconducting leads} 
{\it Phys. Rev.\ B} {\bf 85} 205451 (2011).

\bibitem{Calle-2013}
Calle A M, Pacheco M and Orellana P A 2013
{\rm Fano effect and Andreev bound states in T-shape double quantum dots} 
{\it Phys.\ Lett.\ A} {\bf 377} 1474 

\bibitem{Nozaki-2014}
Nozaki D, Avdoshenko S M, Sevincli H and  Cuniberti G 2014
{\rm Quantum interference in thermoelectric molecular junctions: A toy model perspective} 
{\it J.\ Appl.\ Phys.} {\bf 116} 074308

\bibitem{Trocha-2014}
Trocha P and Barna\'s J 2014
{\rm Spin-polarized Andreev transport influenced by Coulomb repulsion 
through a two-quantum-dot system} 
{\it Phys.\ Rev.\ B} {\bf 89} 245418 

\bibitem{Wojcik-2016}
W\'ojcik K P and  Weymann I 2016
{\rm Thermopower of strongly correlated T-shaped double quantum dots} 
{\it Phys.\ Rev.\ B} {\bf 93} 085428 



\bibitem{Striclet-2012}
Sticlet D, Bena C and Simon P 2012  
{\rm Spin and Majorana polarization in topological superconducting wires}
{\it Phys.\ Rev.\ Lett.} {\bf 108} 096802

\bibitem{Kjaergaard-2012}
Kjaergaard M,  W\"olms K and Flensberg K 2012
{\rm Majorana fermions in superconducting nanowires without spin-orbit coupling}
{\it Phys.\ Rev.\ B} {\bf 85} 020503 

\bibitem{Shi-2016}
Shi Z C, Wang W and Yi X X 2016
{\rm Entangled states of two quantum dots mediated by Majorana fermions}
{\it New J.\ Phys.} {\bf 18} 023005


\bibitem{Vernek-2014}
Vernek E, Penteado P H, Seridonio A C and Egues J C 2014
{\rm Suble leakage of a Majorana mode into a quantum dot}
{\it Phys.\ Rev.\ B} {\bf 89} 165314

\bibitem{Vernek-2015}
Ruiz-Tijerina D A, Vernek E, Dias da Silva L G G V  and Egues J C 2015
{\rm Interaction effects on a Majorana zero mode leaking into a quantum dot}
{\it Phys.\ Rev.\ B} {\bf 91} 115435 


\bibitem{Balatsky-2006}
Balatsky A V,  Vekhter I  and  Zhu J -X 2006 
{\rm Impurity-induced states in conventional and unconventional superconductors} 
{\it Rev.\ Mod.\ Phys.} {\bf 78} 373 

\bibitem{Domanski-2010}
Doma\'nski T 2010 
{\rm Particle-hole mixing driven by the superconducting fluctuations}
{\it Eur.\ Phys.\ J.\ B} {\bf 74} 437

\bibitem{Golub-2015}
 Golub A 2015
{\rm Multiple Andreev reflections in s-wave superconductor-quantum dot-topological
superconductor tunnel junctions and Majorana bound states} 
{\it Phys.\ Rev.\ B} {\bf 91} 205105 


\bibitem{Zitko-2010}
\v{Z}itko R 2011 
{\rm Fano-Kondo effect in side-coupled double quantum dots at finite 
temperatures and the importance of two-stage Kondo screening}
{\it Phys.\ Rev.\ B} {\bf 81} 115316


\bibitem{Bauer-07}
Bauer J,  Oguri A and  Hewson A C 2007
{\rm Spectral properties of locally correlated electrons 
in a Bardeen – Cooper – Schrieffer superconductor}
{\it J.\ Phys.: Condens.\ Matter} {\bf 19} 486211 

\bibitem{Yamada-11}
 Yamada Y, Tanaka  Y and  Kawakami N 2011
{\rm Interplay of Kondo and superconducting correlations in 
the nonequilibrium Andreev transport through a quantum dot} 
{\it Phys. Rev. B} {\bf 84} 075484 

\bibitem{Baranski-2013}
Bara\'nski J and Doma\'nski T 2013 
{\rm In-gap states of a quantum dot coupled between a normal 
and a superconducting lead}
{\it J.\ Phys.: Condens.\ Matter} {\bf 25} 435305 

\bibitem{Fu-2009}
Fu L  and  Kane C L 2009
{\rm Josephson current and noise at a superconductor/quantum-spin-Hall-insulator/super\-conductor junction}
{\it Phys.\ Rev.\ B} {\bf 79} 161408(R)

\bibitem{Rodero-2011}
Mart\'{i}n-Rodero A and Levy-Yeyati A  2011 
{\rm Josephson and Andreev transport through quantum dots}
{\it Adv.\ Phys.} {\bf 60} 899  

\bibitem{Deacon-2010}
Deacon R S,  Tanaka Y, Oiwa A, Sakano R, Yoshida  K, Shibata K, 
 Hirakawa K and Tarucha S (2010)
{\rm Kondo-enhanced Andreev transport in single self-assembled InAs quantum dots 
contacted with normal and superconducting leads}
{\it  Phys.\ Rev.\ B} {\bf 81} 121308(R)


\bibitem{Lohneysen-2012}
H\"ubler F,  Wolf  M J, Scherer T, Wang  D, Beckmann D 
and L\"ohneysen H (2012) 
{\rm Observation of Andreev bound states at spin-active interfaces}
{\it Phys.\ Rev.\ Lett.} {\bf 109} 087004 

\bibitem{Chang-2013}
Chang W,  Manucharyan  V E, Jespersen T S, Nyg{\aa}rd J and Marcus C M (2013) 
{\rm Tunneling spectroscopy of quasiparticle bound states in a spinful Josephson junction} 
{\it Phys.\ Rev.\ Lett.} {\bf 110} 217005

\bibitem{Aguado-2013}
Lee E J H,  Jiang  X, Houzet  M, Aguado  R, Lieber  Ch M and De Franceschi  S (2014)
{\rm Spin-resolved Andreev levels and parity crossings in hybrid superconductor–semiconductor nanostructures}
{\it Nature Nanotechnology} {\bf 9}  79


\end{thebibliography}
\end{document}